\documentclass[aps,prd,amsmath,twocolumn,superscriptaddress,floatfix,nofootinbib]{revtex4}
\usepackage[dvips]{color}
\usepackage{epsfig}
\usepackage{latexsym}
\usepackage{graphicx}
\usepackage{amsfonts}
\usepackage{amsmath}
\usepackage{todonotes}
\usepackage{natbib}
\usepackage{wasysym}
\usepackage{bm}
\usepackage{array}
\usepackage{enumitem} 
\usepackage{color}
\usepackage{xcolor}
\usepackage{blindtext}
\usepackage{ulem}  
\usepackage{comment}
\usepackage{hyperref}
\usepackage{pdfpages}
\newcommand\marina[1]{\textcolor{red}{#1}}
\newcommand\andrew[1]{\textcolor{blue}{#1}}
\newcommand\vasco[1]{\textcolor{magenta}{#1}}
\newcommand{\f}{\begin{equation}}
\newcommand{\ff}{\end{equation}}

\begin{document}

\title{Higher Dimensional Energetic Causal Sets}

\author{Vasco Gil Gomes} 
\affiliation{Departamento de F\'{\i}sica, Universidade de Coimbra, 3004-516, Portugal}
\author{Marina Cort\^es}
\affiliation{Instituto de Astrof\'{\i}sica e Ci\^{e}ncias do Espa\c{c}o, Universidade de Lisboa,
Faculdade de Ci\^{e}ncias, Campo Grande, PT1749-016 Lisboa, Portugal}
\affiliation{Perimeter Institute for Theoretical Physics, 31 Caroline Street North, Waterloo, Ontario N2L 2Y5, Canada}
\author{Andrew R.\ Liddle}
\affiliation{Instituto de Astrof\'{\i}sica e Ci\^{e}ncias do Espa\c{c}o, Universidade de Lisboa,
Faculdade de Ci\^{e}ncias, Campo Grande, PT1749-016 Lisboa, Portugal}
\affiliation{Perimeter Institute for Theoretical Physics, 31 Caroline Street North, Waterloo, Ontario N2L 2Y5, Canada}

\date{\today}

\begin{abstract}
The energetic causal set (ECS) program of Cort\^es and Smolin, whose distinguishing feature is the foundational time irreversibility of the evolution equations of quantum gravity, was initiated ten years ago \cite{ecs}. The model showed the emergence of a time-reversible phase from the time-irreversible foundational regime, but originally only had one spatial dimension. The extension to two and more spatial dimensions has posed a substantial challenge, the higher-dimensional set of solutions having measure zero (requiring infinitely-specified initial conditions). This challenge is overcome here with the extension of the ECS to 2+1 dimensions, invoking a finite interaction cross-section to determine generation of new events and including an adjustable fundamental stochasticity. As in the 1+1 dimensional case, we successfully observe a phase transition into a time-symmetric phase, here through the emergence of crystal-like structures. Due to the irreversible evolution we also witness the discrete dynamical systems behaviour of the 1d+1 case explored in later articles of the ECS program, in which the model is captured in and out of the limit cycles which here take the form of lattice crystals.
In a companion paper we carry out a detailed parameter investigation and study the causal network underlying the set, explaining the emergence of the time-symmetric phase. In Section~\ref{sec:IX} we propose a view of science as a (potential) unifier of threads, and present one such perspective unifying aspects of quantum gravity, biology, and artificial general intelligence (AGI).
\end{abstract}

\maketitle

\tableofcontents                                                  

\section{Introduction}\label{sec:ecs}

Energetic Causal Sets (ECS) \cite{ecs,quantum_ecs,spinfoams, limitcycle, realismI, realismII} are a programme of quantum gravity where space-time is described as a set of events together with a total and a (dynamical) partial causal order where time plays a fundamental role. Time, in the form of an event generator, is responsible for the total order on the set. The operating of the event generator chooses the next instant, which we call a new event. Total order reflects the order by which events are created (birth order) and portrays the true causal order and causal foundations of the programme. At each iteration the event generator makes an evaluation over all available events in the set, according to a rule specified by us, before creating a new event and that makes this a total order.

In quantum gravity proposals, space-time positions typically give origin to particles carrying energy--momentum. In ECS, following the arguments of relative locality \cite{relativelocality}, we take energy and momentum as fundamental, as opposed to emergent. Events live in an energy--momentum space with an assigned Minkowski metric. This is responsible for the partial order on the set that arises from the flow of energy--momentum between causally-related events. In addition to being a fundamental quantity of the model, time is also fundamentally asymmetric. Asymmetric time means the past cannot be reconstructed from a complete knowledge of the present state. In the same way, in ECS models, future states of the system are not fully predictable from a complete knowledge of the present. Numerically, this is achieved via stochastic processes throughout the evolution, making it impossible to completely predict what the next event will be. These two properties of the models namely 
\begin{enumerate}
    \item the fundamental role of (non-emergent) time; 
    \item energy--momenta are taken as fundamental quantities, in opposition to space-time coordinates;
\end{enumerate}
make the ECS programme distinguishable from the causal set program \cite{causalsets,csfd1}.

The ECS programme's ultimate goal is to observe what kind of dynamics are obtained from the fundamental principles that precede the formation of space-time. These were proposed in Ref.~\cite{ecs} and will be reviewed in the next section. In particular we wish to observe whether, out of fundamentally time-irreversible laws, one can obtain approximate time-reversible dynamics. Results of a 1d+1 model have, very promisingly, shown this to be possible, with emergent reversible dynamics being tied to the capture of the system by limit cycles, a common behaviour of discrete dynamical systems \cite{AW,limitcycle}.

The Energetic Causal Set programme is distinct from the Causal Set programme \cite{causalsets,csfd1} in that the causal structure underpinning the emergence of space time is evolving dynamically alongside the emergent space-time manifold. 
In the ECS programme, in addition to the total causal order in the set (a microscopic causal order) created by the event generator, the emergent causal order need not preserve either of the two microscopic causal orders. This phenomena, called disordered causality or discausality, was found in the 1d+1 case and studied extensively in Ref.~\cite{realismII}. This doesn't contradict the idea of irreversible time evolution, since the total causal order representing the fundamental activity of time is never reversed. What happens is that the order by which events are created in the pre-space-time regime need not correspond to the Minkowski order resulting from their embedding in a suitable space-time. In a companion paper \cite{ECS2d2} we will study this phenomenon in the 2d+1 case introduced here.

\section{Energetic Causal Sets}

There are two approaches to introducing irreversibility into a fundamental model. We can have non-bijective laws governing the dynamics, or we can input stochasticity into the laws. Irreversibility in ECS is built numerically into the theory in the form of (pseudo) random inputs in the algorithm determining the dynamics of the set. These dynamics are governed by relativistic energy--momentum conservation laws and new events are created according to a rule that seeks to minimize the differences in the event's causal pasts. The relativistic energy--momentum laws prescribe the flow of energy and momentum between events. Space-time is then to emerge through bulk averages of the causal histories, with the conservation laws acting as constraints on the system as shown in Ref~\cite{ecs}. Note that the time that emerges with space-time, which we denominate {\it Minkowski time} is of a different nature to the fundamental activity of time consisting on creating new events from the events forming the present state of the causal set. 

We start in energy--momentum space labelled by flat Minkowski metric $\eta_{ab}$. ECS models are built up event-by-event from an initial set of events specified by random initial conditions. An event is an exchange of energy--momenta between two (or more) null rays. Different events are connected by these null lines in the same energy--momentum space. The model for the event generator we use here follows the one in the original article \cite{ecs}, where for each event, there are {\it two} incoming momenta and two outgoing momenta. Events are connected by null lines carrying energy--momentum.

Two events (the `parent' events) combine to generate each new one, and each existing event has the capacity to contribute to two new events. Given that each event is the interaction between two null lines, in what follows, for simplicity and algorithmic purposes we will refer to these null lines as {\it particles} or {\it momenta}.

There are four principles that form the foundation of ECS \cite{ecs}:
\begin{enumerate}
	\item Time is a fundamental quantity, an elementary process by which new events are created out of present ones. Causality is a result of this process and its irreversibility, giving rise to a total causal order on the set, the order of birth of events.
	\item Time is unidirectional in the sense that we cannot use the true fundamental laws of the Universe to reconstruct the past out of a complete knowledge of the present, nor can we fully predict the future.
	\item All space-time properties have a dynamical origin, and space-time itself is to emerge from the dynamics. This emergence is responsible for the partial order induced by the metric, flat in our case, and corresponds to the usual Minkowskian causal order.
	\item Energy--momentum is a fundamental property of each event and through laws of conservation dynamics arises. This also gives rise to a dynamical partial order within the set that keeps track of the transfer of energy--momentum, which is controlled by relativistic energy--momentum conservation laws.  
\end{enumerate}

One of the consequences of the third principle is a relational point of view, making each event unique in the causal history of the set. Since each event exists in relation to all others, each event is unique and distinguishable from all others. Hence in our simulations we need to be able to make each event computationally unique. We do that by making use of their causal past, since this is the underlying cause of an event's uniqueness. But causal networks are complex, with an increasing degree of (computational) complexity as more and more events are added, so we need to simplify them in order to obtain a reduced version that is still unique for each event while being computationally feasible. 

This was achieved by storing each newly created event in a subset formed by the causally related  events in its causal past network. We colloquially denote these causal pasts by {\it families} \cite{ecs}. Each initial event defines a family, with lineage inherited through one of the parents. In the 1d+1 case this was taken to be carried by the left-moving ray (it could equally well have been the right-moving one); in 2d+1, where the particles can be moving in any direction, we will need to formalise this by assigning a label to each outgoing ray of the initial events. Since `left' and `right'-moving do not carry through to the 2d+1 case we assign each of the two outgoing momenta a parity-label: `particle' and `anti-particle'. Each momentum preserves this label throughout the evolution of the causal set. For lineage assignment, and in keeping with the original article, we arbitrarily adopt one of these labels for family membership: {\it family membership is carried via the particles, rather than the anti-particles}.

We will implement the algorithm on a compact space with periodic boundary conditions.

In order to characterize the past of each event we need a quantity simple enough to be easily computed, but which retains enough complexity to make each event unique. Following the 1d+1 case the quantity we chose for the 2d+1 case is the averaged space-time distance from the (arbitrary) origin of coordinates of all prior events, in the family of the event in question: 
\f
{\rm past}_I^2=\frac{1}{N}\sum_J(-t_J^2+x_J^2+y_J^2)\,.
\label{pasteq}
\ff
Here $x_J$ and $y_J$ take values in the compact space, and $N$ is the number of events $J$ in the family past of event $I$ plus 1. Given that the right-hand side of Eq.~\ref{pasteq} are valued in the real line, Eq.~\ref{pasteq} ensures that each past is unique while remaining a tractable computation. The difference in the pasts between two events is defined as \cite{ecs}
\begin{equation}
D_{IJ} = \left|  {\rm past}_I^2 - {\rm past}_J^2 \right| \,.
\end{equation}
When two null rays intersect each other a new event is created at the intersection and the family of one of the null rays, the particle, will inherit the new event as its direct descendent (belonging to the same lineage).

We denote by $p^I_{aK}$ the four-momentum incoming to event $I$, coming from event $K$ and $q^L_{aI}$ the momentum outgoing from event $I$ to event $L$, with $a$ the space-time index taking values of 0, 1, and 2 in our 2d+1 case. Then conservation of energy--momentum means 
\f
\it{P^I_a}=\sum_K{p^I_{aK}}-\sum_L{q^L_{aI}}=0\,.
\label{pconservation}
\ff
where the sum $K$ is over the incoming events and $L$ over the outgoing events. We will always take two incoming and two outgoing events.
For simplicity we are adopting a static Minkowski energy--momentum metric, so there are no redshifts. This means that the $p$ and $q$ are equal and do not need to be distinguished: 
\f
p^I_{aK}-q^I_{aK}=0\,.
\label{noredshift}
\ff
If we wish to include expansion of space-time in a cosmological setting, this would be achieved by the introduction of an expansion/contraction function on the right-hand side of Eq.~(\ref{noredshift}). 

Finally, since we are dealing with null rays, both energy and momenta are constrained by the usual relations:
\f
\eta^{ab}p^I_{aK}p^I_{bK}=0,  \;\;  \eta^{ab}q^I_{aK}q^I_{bK}=0\,.
\label{eprelations}
\ff
Remember there is no space-time and events live in energy--momentum space, so $\eta^{ab}$ is the metric in this space, which naturally is taken to be the Minkowski metric. The system is set up in such a way that all these equations are automatically satisfied. Null rays are chosen for this reason, since when they interact they can be taken to cross each other without altering their momenta. 

\subsection{ECS in 1+1 dimensions}\label{sec:1decs}

The 2d+1 system we will describe is an extension of the 1d+1 system of Ref.~\cite{ecs}, which we quickly recap. In the 1d+1 case the particles can be taken as point-like, as they will inevitably collide when they cross. We adopt periodic boundary conditions
\f
(x+L,t)\equiv(x,t)\,.
\label{boundaryconditions1d}
\ff
Energy--momentum transfers are constrained by Eqs.~(\ref{pconservation}) and (\ref{eprelations}) and the metric in energy--momentum space is taken to be flat, so no redshifts are permitted, Eq.~(\ref{noredshift}).

\begin{figure*}	
\centering	
\includegraphics[width=0.8\textwidth]{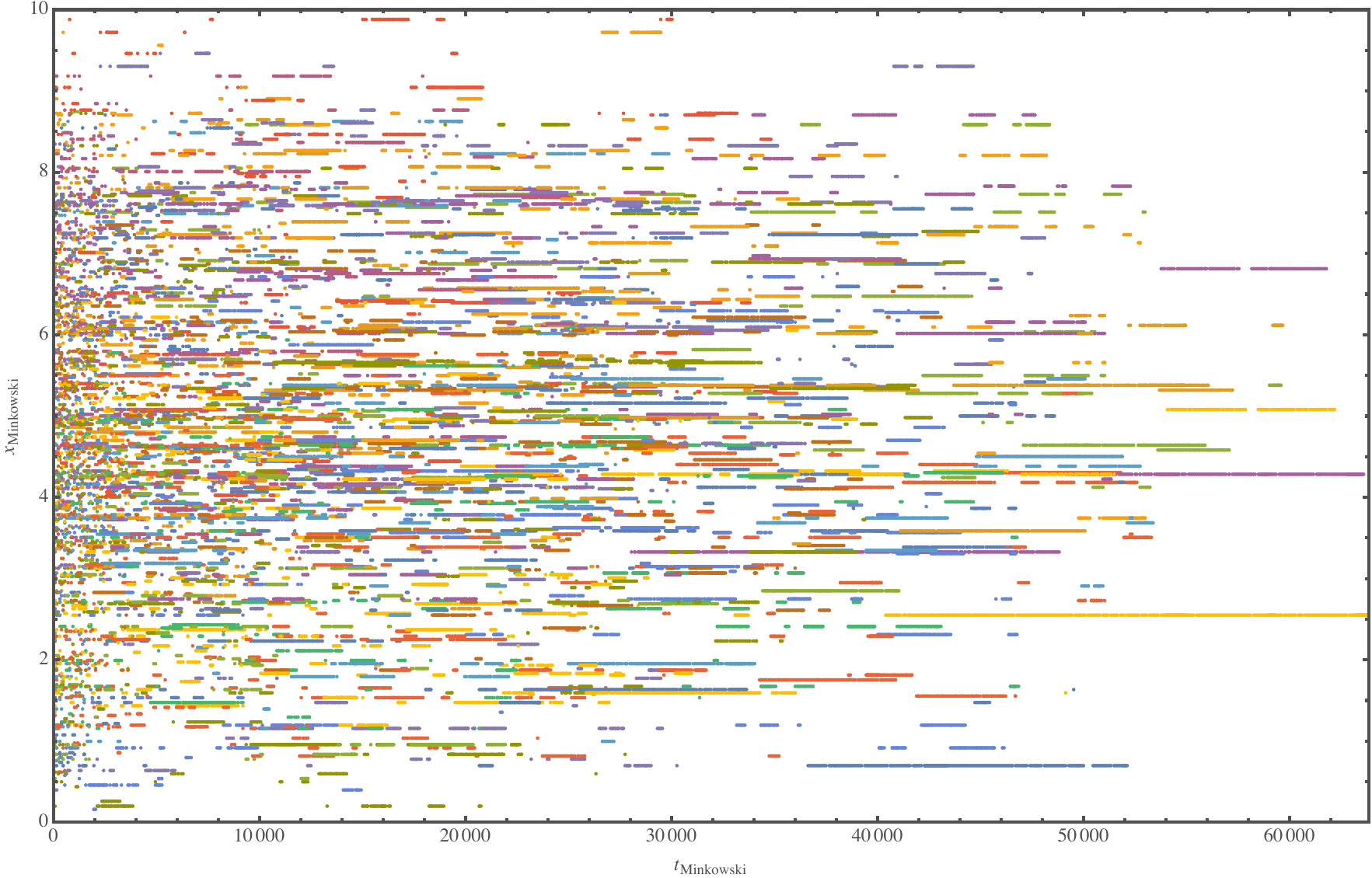} 	
\caption{A run with 40 families and $5\times 10^4$ generated events in 1d+1. Each color represents a family and each dot an event. Initially there is a seemingly chaotic behaviour where all pasts interact, being the time-irreversible phase of the run. Later a structure of stable trajectories arises, characterizing the time-symmetric phase.} \label{fig:1d_quasiparticles}
\end{figure*}

In the 1d+1 system there are left-moving  and right-moving null rays. As they all travel at the speed of light, only oppositely-moving pairs can interact. The interactions are easy to compute once the two rays have been chosen to create a new event. Parent events are chosen by a rule that seeks to minimize the difference in pasts, which in the 1d+1 case is just
\f
{\rm past}_I^2=\frac{1}{N}\sum_J(-t_J^2+x_J^2)\,.
\ff
The coordinates of the new event are then given by \cite{ecs}
\f
\begin{cases}
t_{\rm new}=\frac{1}{2}(t_1+t_2\pm x_1\mp x_2+(n-m)L)\\
x_{\rm new}=\frac{1}{2}(\pm t_1\mp t_2+x_1+x_2+(n+m)L)\\
\end{cases}
\ff
where $x_1<x_2$ and upper and lower $\pm,\mp$ refer to right- and left-moving particles respectively. The variables $n$ and $m$ are the number of windings around the compact space each null ray undergoes before interaction. Due to periodicity, there are an infinite number of intersections on the continuation of their null lines. The possible intersections specified by $(n,m)$ is \cite{ecs}
\f
(n,m)=\{(n,n),(n,n+1),(n+1,n)\} \,,
\ff
the two latter cases corresponding to the right- or left-moving ray crossing the boundary first. Picking the closest intersection 
reduces the system to one of simple oscillators with particles interacting only with the ones to its right or left. A stochastic input is introduced in the model by randomly taking $n=\{0,1\}$ in the generation of new events. Once an event is created it is then stored in the family of the left-moving null ray (with equivalent dynamics if choosing the right-moving null ray instead). 

The results of the 1d+1 case show a phase transition from a chaotic time-irreversible phase to a more organized time-symmetric phase, as demonstrated in Ref.~\cite{ecs} and illustrated by Fig.~\ref{fig:1d_quasiparticles}. After enough events have been created, the pasts have built sufficient information that only a few will be selected by the event generator to interact and create new events. In the beginning of the run all families are equivalent. When the creation of new events begins pasts start to separate from each other and as the runs progresses only those that are closer continue to be selected to interact, making them yet closer with each interaction. This is because the pasts eventually become time dominated; this is the meaning of `pasts have built sufficient information'. When this happens stable trajectories, characterized by the same spatial position, appear. These are the 1d+1 quasi-particles that mark the emergence of the time-symmetric phase of the evolution. When this happens novelty, in the form of the various available spatial positions for the creation of new events, has been exhausted with only a few families interacting and with new events being born in only a small number of different spatial positions. 

\subsection{ECS in 2+1 dimensions}\label{sec:2decs}

In this section we will describe how we solved the main obstacle that hitherto prevented extension of the ECS programme towards higher dimensions. 
We envisage a system of events linked by null lines. For algorithmic ease we associate with each ray a number, loosely termed `parity-label', for consistent tracking and record-keeping of each ray in the causal network. We only allow rays with different parity-labels to interact. Two momenta going out from the same event are additionally not permitted to generate a second event. 

In the 2d+1 case, just like in the 1d+1, there needs to be enough diversity (present both in the initial conditions and in the event generator algorithm) to ensure that the desired phase transition to the time-symmetric phase doesn't occur promptly when the simulation starts, leading to a trivial model of no interest for our purpose. We shall use interchangeably diversity, novelty, and genetic pool. The last capitalizes on the analogue of evolutionary theory: the need for sufficient diversity in the genetic pool of existing elements to ensure adaptability of the surviving lineage, and prevent the so-called inbreeding problem of living systems.

Each event is generated by the intersection of two null-rays labelled `particle' and `anti-particle', so each event has two incoming and two outgoing momenta. The present is defined to be the set of events whose outgoing null rays have not yet both been utilized. This is sometimes referred to as the {\bf thick present} as we have called in previous articles in the ECS program. The past is the set of events whose null rays have already both been utilised. We are working in a compact space with boundary conditions analogous to the 1d+1 case, so our space is a square of length $L$. Particles therefore cross one side and return by the opposite side.  Of course the ideal is to make the compactification scale as large as possible within computational constraints, to obtain results which become independent of the compactification scale, as was already shown for 1d+1 in Ref.~\cite{ecs} through simulations in extended space. 

We start our simulations at global time $t_{\rm Minkowski}=0$ by randomly spraying the total number of chosen initial events into our 2d compact space and randomly choosing the azimuthal angles $\phi$ defining the direction of the particles. Taking $c=1$ our particles will have $45^\circ$ angles with the time axis of the Minkowski space-time. The directions of the particles and anti-particles in 2d space are chosen randomly. 

The number of families gives the number of causal pasts to track. At each step a particle and antiparticle are chosen to interact, either at random or as those which minimize the difference in the pasts amongst their parent events (i.e.\ all events belonging to the thick present). Their future collision's space-time coordinates are computed with the two incoming momenta defining the new event.

The key issue in extending the simulations to a 2d+1 or higher-dimensional version, and herein lies the crux of our work, is how to mathematically define the interactions in a self-consistent manner. In higher spatial dimensions, such as 2d+1, particles cannot be treated like points as there will be no collisions since the set of intersecting lines has zero measure even in the compactified space. To ensure interactions, the particles must therefore be given a finite extent, or cross-section, which will be an additional fundamental parameter defining the model. Coupled with the need to maintain family number as identified above, these interactions should be between a particle and an anti-particle, as in a normal quantum field theory. 

Giving particles a 2d extent implies that their trajectories trace out tubes through space-time. Having used the pasts to identify the particle/anti-particle pair to collide, our first task is to compute when and where the collision takes place. The centres of each tube are given by parametric equations, and the collision occurs when the two centres first pass within a distance of twice the radius. To identify this we step along the trajectories with a timestep $dt$, which must be chosen small enough that collisions are not accidentally missed.\footnote{There may be an opportunity to streamline this search using a Monte Carlo minimising algorithm, which we defer to future work.}

Unlike in 1d+1, where particles always collide head-on at a point, the typical 2d+1 collision has a finite closest distance of the centres, opening the question of how to define the location of the new event that the collision generates, and how the new null rays emerge from it. For instance, should their emission location be shifted to the centre-of-mass of the original collision? This is key because in the 1d case one of the features that caused the exhaustion of space-time locations, allowing quasi-particles to emerge, was the fact that not only does the set of initial positions determine all future spatial collision locations, but also that this set is finite. Anything that shifts the position of the particles during collisions amounts to effectively changing the initial conditions, which in turn is equivalent to constantly injecting additional randomness that the system cannot overcome. We will discuss in Section~\ref{sec:intersection} how we resolved this.

\section{2+1 dimensional model}

Our new 2d+1 simulations, as in the 1d+1 case, were carried out using the \texttt{Mathematica} environment. We will now describe the algorithm and parameters controlling the runs. We will explain how we define intersections and then the general layout of the code.

\subsection{Parameters}

We first consider the parameters controlling the runs. We are working in a compact 2d space which for simplicity we take to be a square of side $L$.

Since particles have a 2d extent we need to define their radius. In practice it is not the absolute radius which is relevant, but its size in proportion to that of our compact space. We define the ratio of the radius and the side of the square as the variable controlling the size of our particles and anti-particles. We called this variable the `cross-section', $\bar{\sigma}$. 
The radius of a particle is  $\bar{\sigma}\times L$.

Once we identify the particle/anti-particle pair to collide based on their pasts, we impose a limit on how long we will wait for them to collide, $t_{{\rm max}}$, in analogy with the 1d+1 case where particles was not allowed to undergo more than two windings before a new event was created.  Comparing $t_{\rm max}$ to the length of our square estimates how many windings round the compact space the particles can travel before interacting. For example $t_{{\rm max}}=L$ allows only roughly one winding for an interaction to occur. If the chosen pair fails to interact within the allowed interval, we switch to the pair whose difference in pasts is the second lowest, then the third and so on until we find a pair that collides.

We can get insight into suitable values of $t_{\rm max}$ by calculating the mean time to collision, averaged over trajectory directions. Moving to the rest frame of the anti-particle, the particle will have an effective average r.m.s.\ velocity of $\sqrt{2}$, and given its radius $\bar{\sigma} L$ will trace out an area $2\sqrt{2}\bar{\sigma}L t$ in a time $t$. The mean time to collision $\bar{t}$ is when this area reaches half the total spatial area $L^2$, i.e.\
\begin{equation}
\label{e:tbar}
\bar{t} = \frac{L}{4\sqrt{2} \bar{\sigma}} \,.
\end{equation}
To give the chosen particles a good chance to collide, $t_{\rm max}$ should be significantly greater than this. 

Finally we have a parameter, $\Omega$, which is the amount of stochasticity we give the run, see Figure~\ref{fig:main_result} for an example where $\Omega=0.001$. Randomness is introduced by having, with some probability, the parents being randomly selected rather than according to our past-difference rule. The longer it takes to transition to the time-symmetric phase without randomness, the less randomness we can allow. To tune this parameter we need first to study the runs without randomness to estimate how many events are needed to enter a time-symmetric phase for chosen parameters. For instance, suppose it takes around 1000 events to do so. Given that one randomly-generated event can be enough to disrupt the time-symmetric phase, as we will discuss in the Appendix, we should then consider how many random events we want per 1000 events. 

\subsection{Defining an intersection}\label{sec:intersection}

Having computed the intersections, which we do by solving the parametric equations for the selected rays with a suitable time step $dt$,  we must specify what coordinates to give to the new event. As we discussed in Section~\ref{sec:2decs}, in the 1d+1 case the initial conditions fully determine every possible collision between any two pairs, determining the full finite and fixed set of possible spatial locations for all future interactions. In the 1d+1 case the particles can be taken to be points and defining coordinates for the new event is simple. Moreover simple expressions that determine all future interactions between any oppositely-moving pair can be obtained. 

This is not the case in 2d+1. The particles cannot be taken as point-like as they would never collide, but having taken them as extended objects the collisions too become extended regions in space-time. Despite this we still want to define events as space-time points.\footnote{Issues regarding the definition of simultaneity in extended regions of space-time will arise if we try to work with some sort of extended definition of the collision point \cite{becoming}.} What point should we define to be the coordinates for the new event? 

The most natural candidate is the midpoint between the centers of the particles. But if at each new event we redefine the position of the rays that gave birth to it by shifting them to a midpoint, the effect is like changing the initial position of that ray, therefore adding new coordinates to the initial conditions. Remember that both rays are supposed to leave the newly-created event and that the initial conditions determine all future possible spatial positions for collisions. In the 1d+1 case it is the fact that there is a fixed number of initial conditions that allows some of the spatial positions to be exhausted and the phase transition to occur in the form of emergent quasi-particles. In the 2d+1 model, shifting the collision coordinates to a midpoint would constantly add new initial conditions, so that when creating a new event we are in practice restarting the run all over again. The result would be that the phase transition can never occur. 

The resolution was to define the coordinates of the new event to be the ones of the particle at the time of collision. This means that the anti-particles will be shifted at each interaction. This will not be an issue, since we are assigning the new event to the family of the particle, so that computationally particles are the determining factor here. Naturally the same results would be replicated with the anti-particles by storing the new events in the families of the anti-particles. This choice then allowed the time-symmetric phase to emerge. 

\subsection{Overview of the algorithm}\label{sec:2d_model}

We start our simulations by choosing the parameter values, which define the model and specify the algorithm details. The physical parameters governing the model are the periodicity of space $L$, the number of families $N_{\rm families}$, the particle interaction cross-section $\bar{\sigma}$, and the stochasticity $\Omega$. The algorithm is controlled by the number of events chosen to be simulated $N_{\rm events}$, the collision search resolution $dt$, and the maximum collision search time $t_{\rm max}$. We then give random initial energy-momenta values to the events and random azimuthal angles to our so-called particles and anti-particles.

\begin{figure*}%[htpb]	width=80mm,height=55mm
	\centering
	\includegraphics[width=0.9\textwidth]{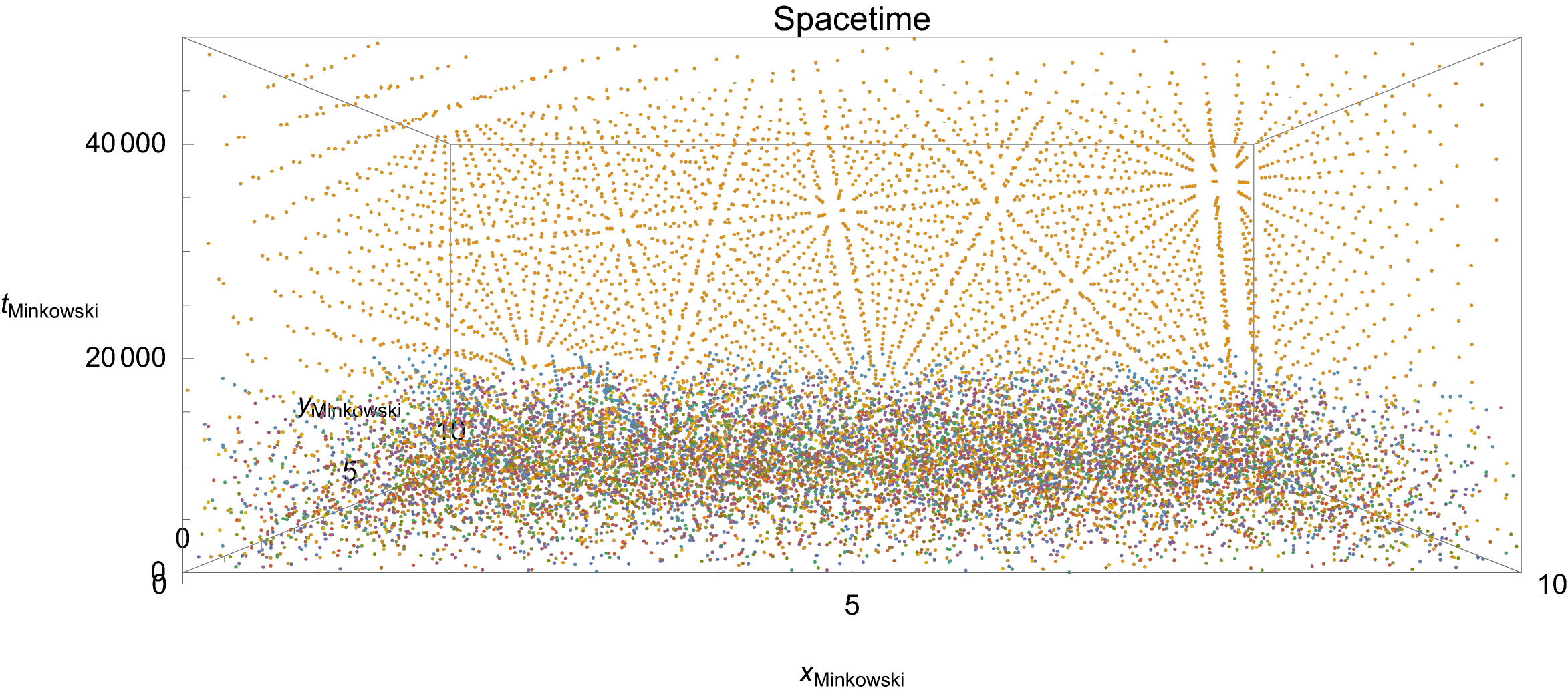} \caption{A deterministic run with $N_{\rm families} = 30$ and $N_{\rm events} = 3.5 \times 10^4$. Parameters: $L=10$, $\bar{\sigma}=0.07$, $dt=\bar{\sigma}$, $t_{{\rm max}}= 3 L$. Each color represents a family and each dot an event. The run starts with a seemingly chaotic behaviour where all families interact, then a structure of stable trajectories resembling a crystal emerges. This run is presented without any randomness being injected into system, aside from the generation of initial conditions. We see the emergence to the time-symmetric phase characterized by the ordered crystal structure, unlike the time-asymmetric phase where are families are interacting and the system can get to any of its limit cycles.  Figure~\ref{fig:deterministic_crystal_views}, shows two different viewing angles of the same crystal.} 
	\label{fig:2d_quasiparticles}
\end{figure*}

To begin the evolution, we then choose a particle/anti-particle pair to interact. First a random number between 0 and 1 is generated; if this number is smaller than the chosen randomness $\Omega$ the interacting pair will be chosen at random (making sure the particle and anti-particle don't belong to the same event), repeatedly so if the particles do not interact within $t_{\rm max}$. Otherwise, we use the deterministic rule of closest pasts. The difference between the pasts of the parent events of all particle/anti-particle pairs is made, except those coming from the same event.  If the closest pair doesn't collide within time $t_{{\rm max}}$ the second closest is selected and so on. If we fail with all pairs the run is aborted. Appropriately chosen algorithmic parameters will avoid this.

\begin{figure*}%[t]
	\centering
$\begin{array}{cc}		
	\includegraphics[width=120mm,height=66mm]{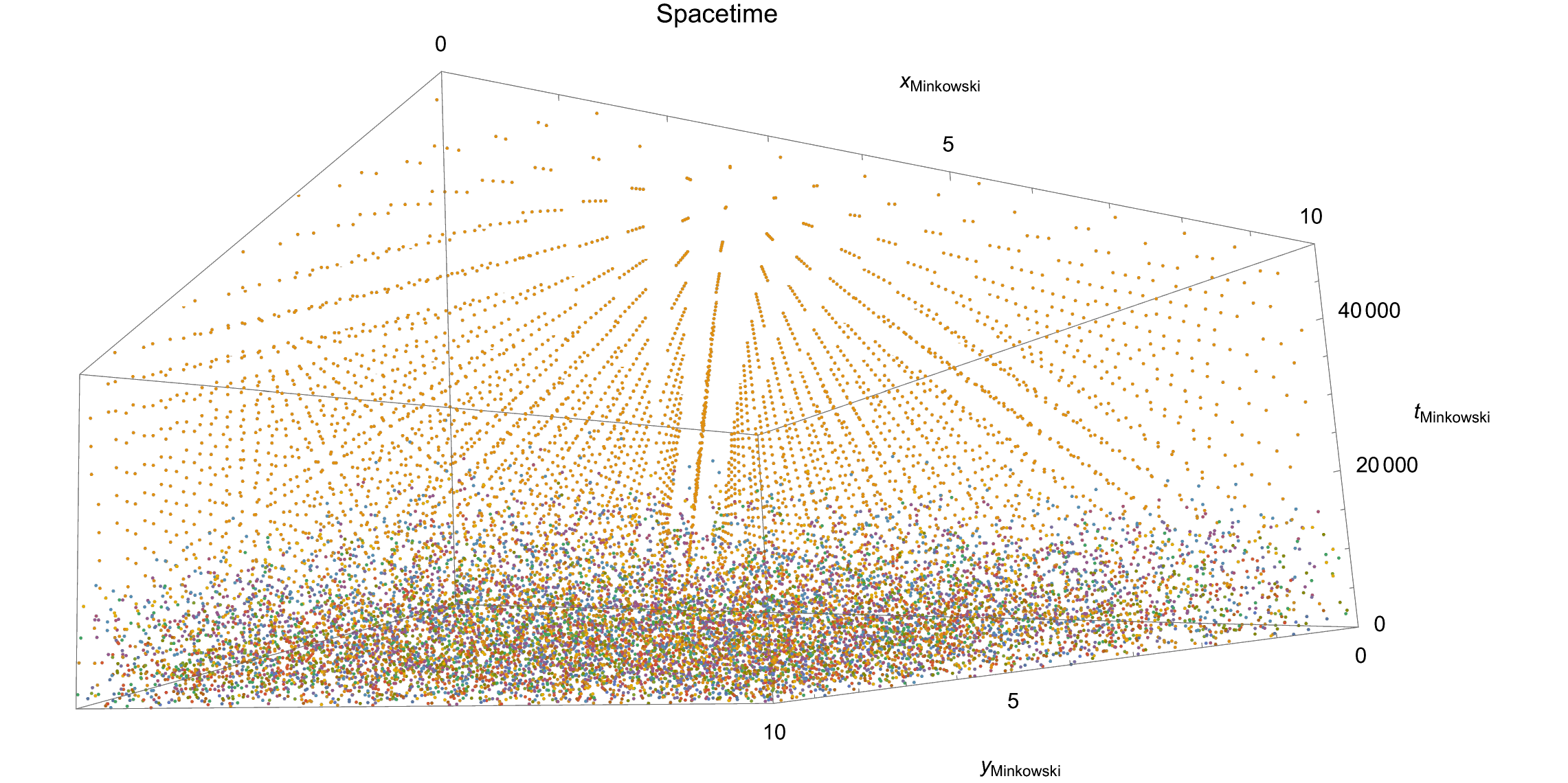}\\
    \textbf{(a)} \\
    ~\\
	\includegraphics[width=120mm,height=66mm]{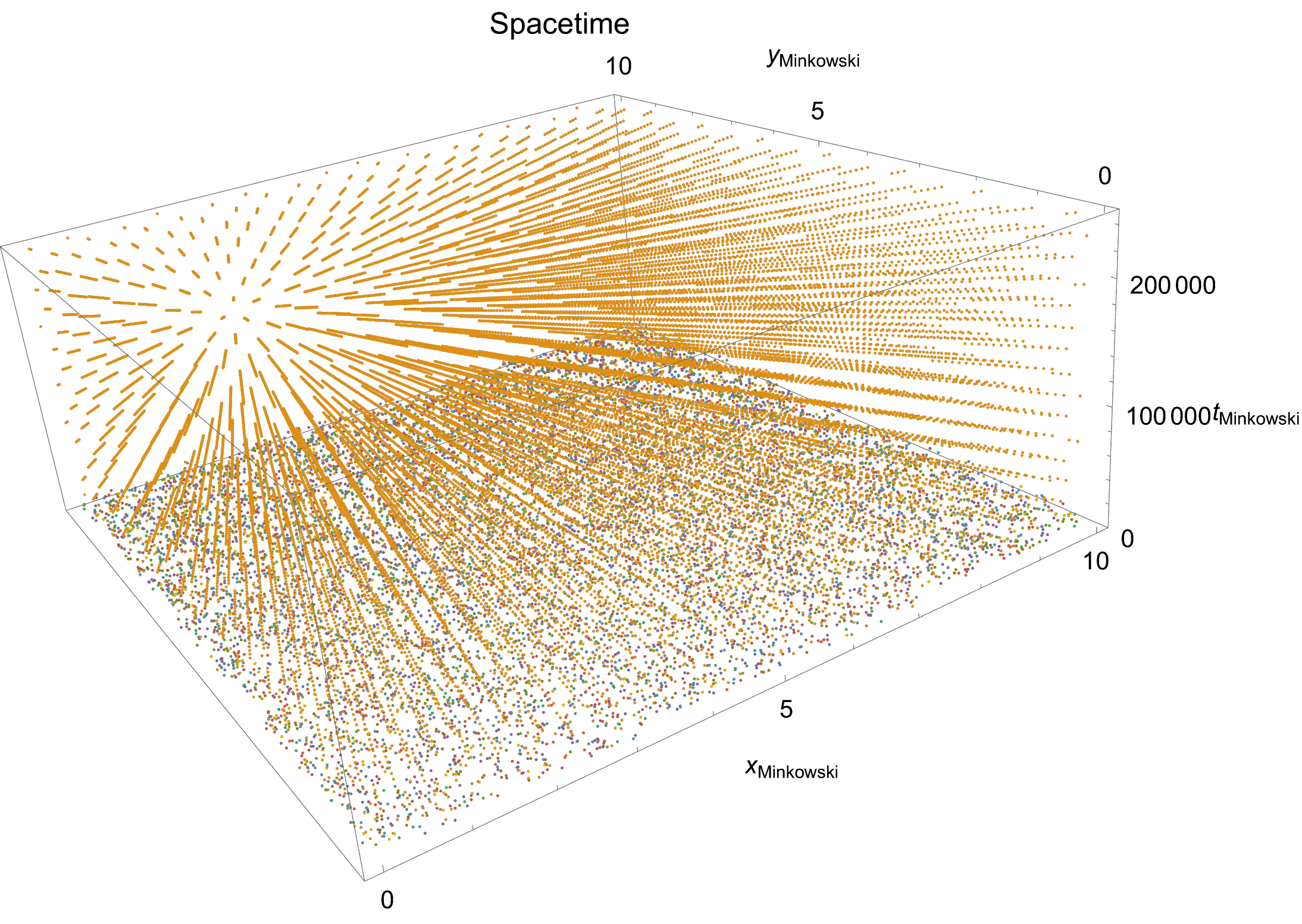}\\
    \textbf{(b)} \\
	\end{array}$
	\caption{Two distinct viewpoints (angles of view) of Figure~\ref{fig:2d_quasiparticles}. } 
	\label{fig:deterministic_crystal_views} 
\end{figure*}

To ensure the new event is created after its parents the collision search begins at the birth time of the younger parent. Having created the new event, we update the present and the pasts before proceeding to the next event. The events are stored in a table organized by families. As in the 1d+1 case, events interact twice (one particle and one anti-particle available for each). Parameter tuning is important for the efficiency of each run.

The procedure can be summarised as follows:
\begin{itemize}
    \item \textbf{Step 0:} Initialize.\\
    Choose the total number of events to be generated $N_{\rm events}$. Pick $N_{\rm families}$  initial events at random spatial positions and initial global time $t_{\rm Minkowski}=0$. Give their emitted particles and anti-particles random azimuthal angles $\phi$. Select the cross-section $\bar\sigma$ of the particles, and the randomness $\Omega$ for the run. Finally, choose suitable algorithmic parameters $dt$ and $t_{{\rm max}}$.
    \item \textbf{Step 1:} Create a new event.\\
    Pick a particle and anti-particle (randomly or by minimizing the differences in the pasts of their parent events $D_{IJ}$) able to interact within the chosen $t_{{\rm max}}$ and construct the new event to the future of $I$ and $J$.
    Select the particle from event $I$ and the anti-particle from event $J$. Because we have $2N_{\rm families}$ particles, half of them anti-particles and half particles, and since particles from the same family can't interact, we have $N_{\rm families}(N_{\rm families} - 1)$ possible pairs. For example pairs $1,2$ and $2,1$ would mean particle from event $1$ and anti-particle from event $2$ in the first case and the opposite in the second. 
    \item \textbf{Step 2:} Compute the past of the new event and store it in the family of the parent that contributed the particle. Remove the particle and anti-particle from their respective sets. This is to make sure a parent event that has used both its particle and anti-particle is removed from the thick present.
    \item \textbf{Iterate:} Return to Step 1 and repeat $N_{\rm events}$ times.
\end{itemize}

\subsection{2d quasi-particles}

We finish this section by presenting a plot with the sought-after phase transition, which in the ECS programme shows how one obtains approximates time-symmetric dynamics from foundational time asymmetry in the quantum gravitational regime. As in the 1d+1 case all simulations are characterized by two phases, a seemingly chaotic and disordered phase and an ordered phase. The latter is the lock-in phase, where stable space-time trajectories, quasi-particles, are formed. In 2d+1 these have the visual appearance of a crystal. The disordered phase is the time-asymmetric phase of the event generator, manifesting itself through lack of a pattern in the space-time positions of the events. As the system enters the lock-in phase, time symmetry emerges in the form of stable trajectories. In the disordered phase all family sets partake in the generation of new events, but as the lock-in phase settles in only a few of them will keep interacting. 

Figure~\ref{fig:2d_quasiparticles} shows this. There is no randomness in this run; all events are generated deterministically by the event generator, which preserves the crystal formation after the phase transition. There are 40 families and a total of $5 \times 10^4$ events, with each color representing a different family. When the time-symmetric phase emerges there is an interlocking between the same particles and anti-particles which produces the pattern. Given that we have a one-color crystal we interpret this interlocking to be between one particle and two anti-particles (it has to be two anti-particles, since we don't allow a pair of the same event to produce a new event). Figure~\ref{fig:deterministic_crystal_views} shows the structure from two different viewing angles.

Even though Fig.~\ref{fig:2d_quasiparticles} shows a simulation with a fully-deterministic event generator, there is a degree of randomness in every run from the randomly-generated initial conditions. The system needs to overcome this initial randomness, which makes every past eligible for interaction, entering a more ordered deterministic phase where pasts have built enough information that only a few remain eligible for interaction.

We will see in the next section how some of the parameters influence the emergence of the crystal, leaving a more detailed study for the Appendix. In a companion paper~\cite{ECS2d2} we will delve deeper into how these structures are forming by analysing the underlying causal structure of the set.

\begin{figure*}%[t]
	\centering
$\begin{array}{cc}		
	\includegraphics[width=80mm]{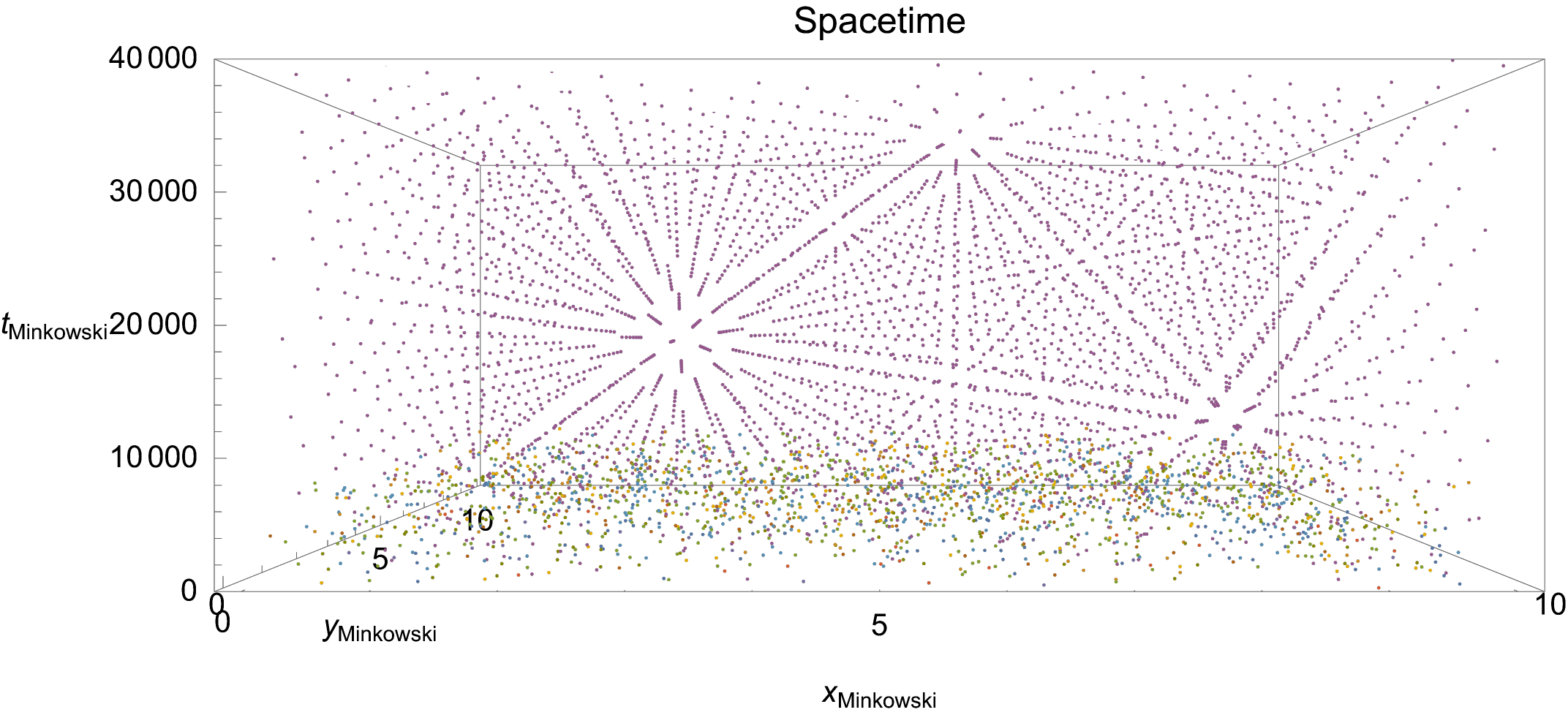} \quad &\quad 
	\includegraphics[width=80mm]{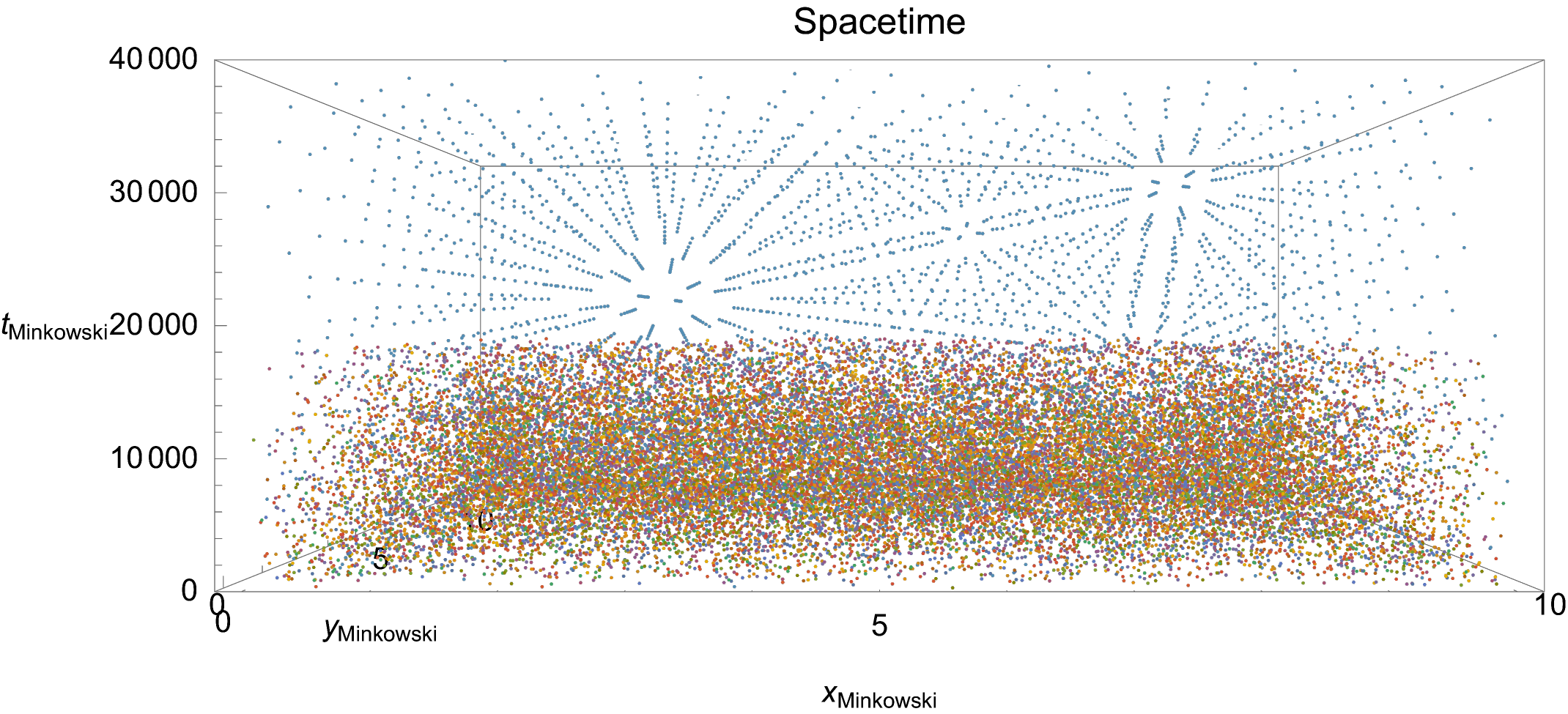}\\
		\textbf{(a)} & \textbf{(b)} \\
		~\\
		\end{array}$
	\caption{The effect of the number of families on runs with zero randomness. We choose zero randomness for these tests to highlight the effect of the parameters without adding other effects into the mix. Parameters: $L=10$, $dt=\bar{\sigma}$, $t_{{\rm max}}=2\times L$, $\bar{\sigma}=0.07$, $N_{\rm events} = 5 \times 10^4$. Plots \textbf{(a)} and \textbf{(b)} show $N_{\rm familes} = 10$ and 50 respectively. As we increase the radius the crystal forms more quickly, as the number of families we are increasing in effect the number of initial events. Since these are generated randomly we are increasing the initial random input the system needs to overcome before it can settle in a limit cycle and form a crystal. Therefore as the number of families increases the longer it will take for the transition to occur.}
	\label{compare_fam} 
\end{figure*}

\section{Parameter control and system tuning}

In this section we will quickly describe the effect of varying the physical parameters $N_{\rm families}$ and $\bar\sigma$, before moving on to Section~\ref{sec:randomness} where we add the final detail to the simulations by introducing numerical randomness to the dynamics. The following figures show fully deterministic event generators, to show the influence of the parameters on the outcome and how they relate to each other, so that we can efficiently tune the runs before adding stochasticity. In the following plots, each dot represents an event and each color represents a family.

\subsection{Number of families: \protect{$N_{\rm families}$}}
\label{sec:compare_families}

At the beginning of each run there is random input in the form of the initial conditions; the spatial positions of the events and 2d trajectory angles of the particles. Before a time-symmetric phase can arise this randomness, responsible for disorder and time asymmetry, needs to be overcome. The larger the number of initial events/families the larger the random input, making it take longer for the system to overcome the initial irreversible phase and reach one of its limit cycles \cite{limitcycle}. The expected behaviour is therefore that as the number of families increases, the phase transition will take more time to occur. A bigger number of events happening usually results in a longer Minkowski time, but this relation isn't strict as events are often created with an earlier Minkowski time than some of their predecessors (though never before their parents!).

Figure~\ref{compare_fam} shows this phenomenon. Plot \textbf{(a)} shows a run with $5 \times 10^4$ events and 10 families and while plot \textbf{(b)} has the same number of generated events but 50 families. The differences are clear, with only 10 families the transition occurs almost promptly whereas with 50 families it took a while before the transition could occur and a lot more events had to be generated before the crystal could form.\footnote{In fact with 10 families usually the transition happens even faster and with 50 families usually it takes even a bit longer, so that in general the differences are even bigger.} The runs are truncated at $t_{\rm Minkowski} = 4\times 10^4$ so that phase transition is more clear. Adding another position and an angle to the set of initial conditions of the 1d+1 case had quite a drastic effect in the system, since in the 1d+1 case 7\,500 events were usually more than enough for the emergence of a time-symmetric phase in the 1d+1 case for the number of families presented here.

\subsection{Cross-section: $\bar\sigma$}\label{sec:comparecrazy}

This parameter determines the spatial size of a particle. It is the ratio between the particle radius and the length our square. A small radius, desirable for a classical limit, requires a bigger $t_{{\rm max}}$ in combination with a bigger number of families which affects the length of the run and the speed of convergence to a time-symmetric phase. Moreover we choose $dt$ to be equal to $\bar\sigma$ which also influences the length of the run since it will require more computational power to compute one intersection between two rays.
We will see in the Appendix that $N_{\rm families}$, $\bar\sigma$ and $t_{{\rm max}}$ are three interconnected parameters.

\begin{figure*}%[t]
	\centering
$\begin{array}{cc}		
	\includegraphics[width=80mm]{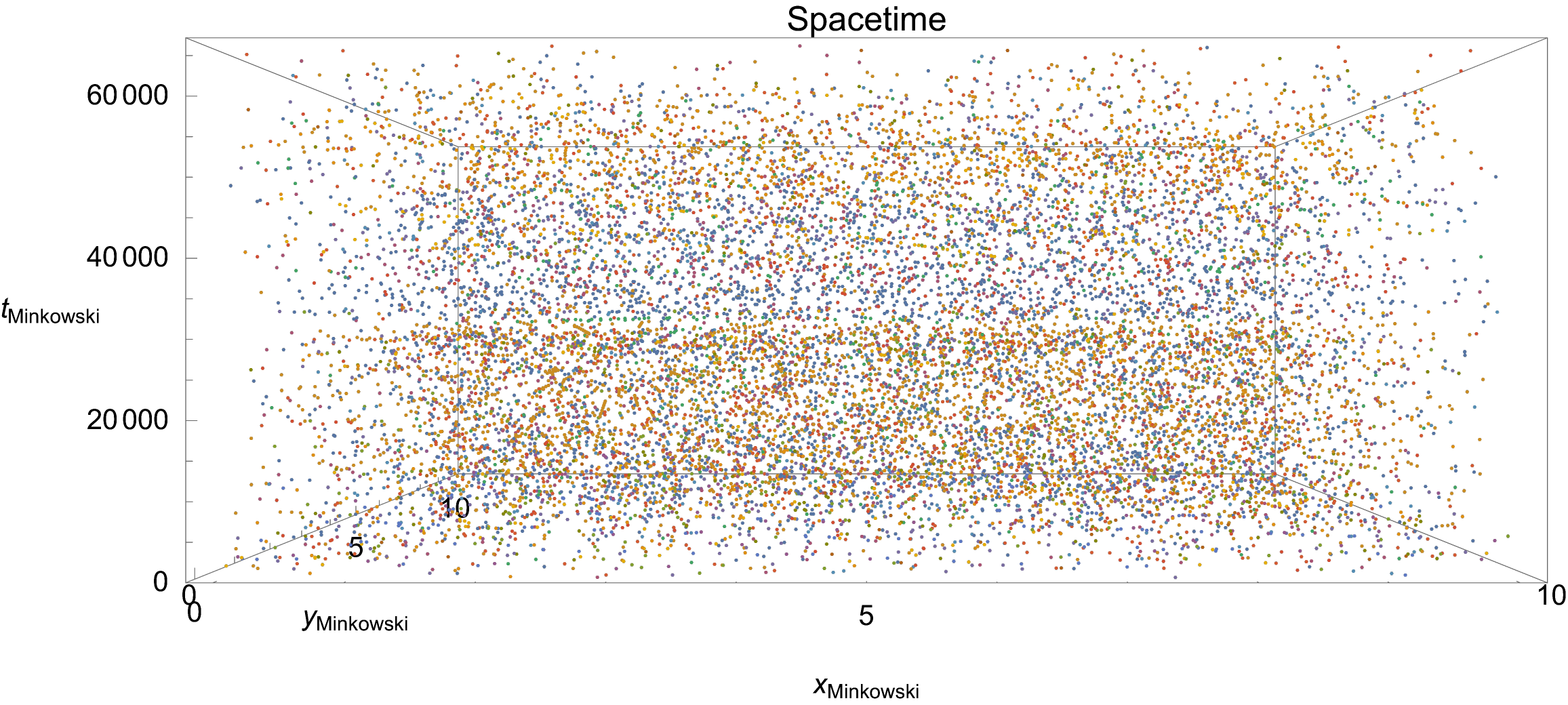} \quad &\quad 
	\includegraphics[width=80mm]{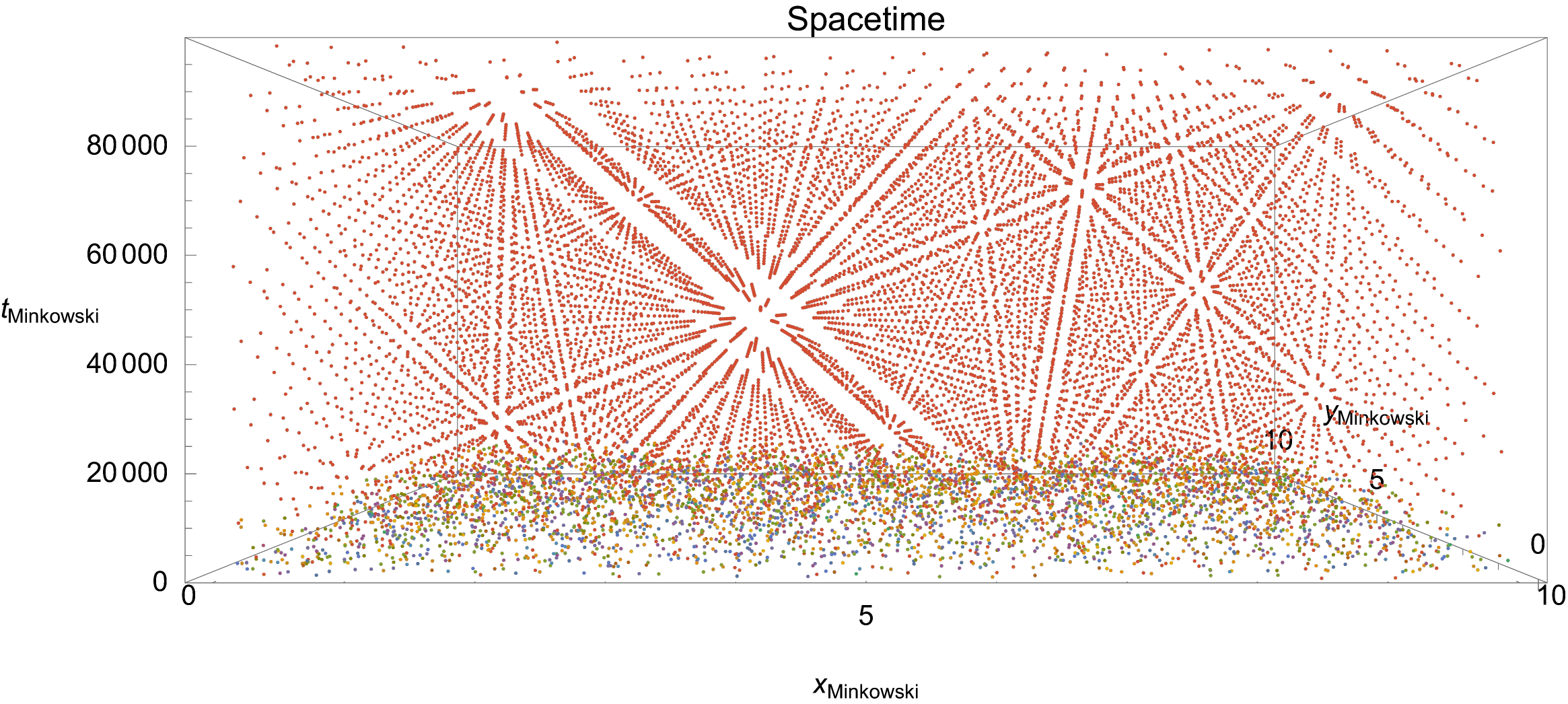}\\
		\textbf{(a)} & \textbf{(b)} \\
		~\\
		\end{array}$
	\caption{The effect of $\bar{\sigma}$ on runs with zero randomness, $N_{\rm families} = 20$ and $N_{\rm events} = 1.5 \times 10^4$. %We choose zero randomness for these tests to highlight the effect of the parameters without adding other effects into the mix. 
 Parameters: $L=10$, $dt=\bar{\sigma}$, $t_{{\rm max}}=2\times L$. Plot \textbf{(a)} has $\bar{\sigma}=0.02$ and plot \textbf{(b)} had $\bar{\sigma}=0.05$. As we increase the radius of the rays collisions will happen more promptly. As the radius becomes smaller, as in plot \textbf{(a)}, the run need to be lengthier in order for the system to settle in a limit cycle. In this case the run wasn't long enough for a crystal to form. A more detailed explanation can be found in the Appendix.}
	\label{compare_crazy} 
\end{figure*}

Figure~\ref{compare_crazy} shows two runs with $N_{\rm families}=20$ and $N_{\rm events}=1.5 \times 10^4$ generated events. We only vary $\bar\sigma$ from 0.02 to 0.05. We can see in plot \textbf{(a)} that the run wasn't long enough for the crystal to emerge, either that or we should have increased $t_{{\rm max}}$, while in plot \textbf{(b)} emergence happened. If the cross-section becomes too big, most pairs will already overlap at the time of birth of the selected youngest parent. This will make the code look for suitable pairs until it finds one that isn't in this condition, spoiling our rule. We had to avoid allowing already intersecting pairs at the time of birth of the youngest parent to be used for the creation of a new event since this would lead to several bugs. It can also happen that all particles are overlapped and there will be no pair available and the run will terminate; we must avoid this uninteresting regime. As the cross-section increases it becomes easier for the time-symmetric phase to emerge since rays will intersect faster.

\section{Introducing irreversibility: $\Omega$}

\label{sec:randomness}
In this section we add the final ingredient that completes the full simulation of energetic causal set models in 2d+1: the introduction of stochasticity in the evolution of the system, which represents the time-irreversibility of dynamics in the quantum gravity description.
Stochasticity will be added via the parameter $\Omega$, which represents the ratio of number of events generated by randomly selection of the pair of parents, $N_{\rm events,rand}$ to the number of events selected by using the event generator, $N_{\rm events,gen}$. 
\f
\Omega=\frac{N_{\rm events,rand}}{N_{\rm events,gen}}
\ff
All models generated up until this section had all events created by the event generator, $N_{\rm total}=N_{\rm events,gen}$.

In the 1d+1 case at each iteration there were constant inputs of randomness introduced by randomly selecting how many windings particles would make before interacting. This was straightforward since within each winding every pair meets once. In the 2d+1 case it is not so simple because particles have an extent and will be interacting for an extended region of space-time, not a point. 

\begin{figure*}%[htpb]	width=80mm,height=55mm
	\centering
	\includegraphics[width=1\textwidth]{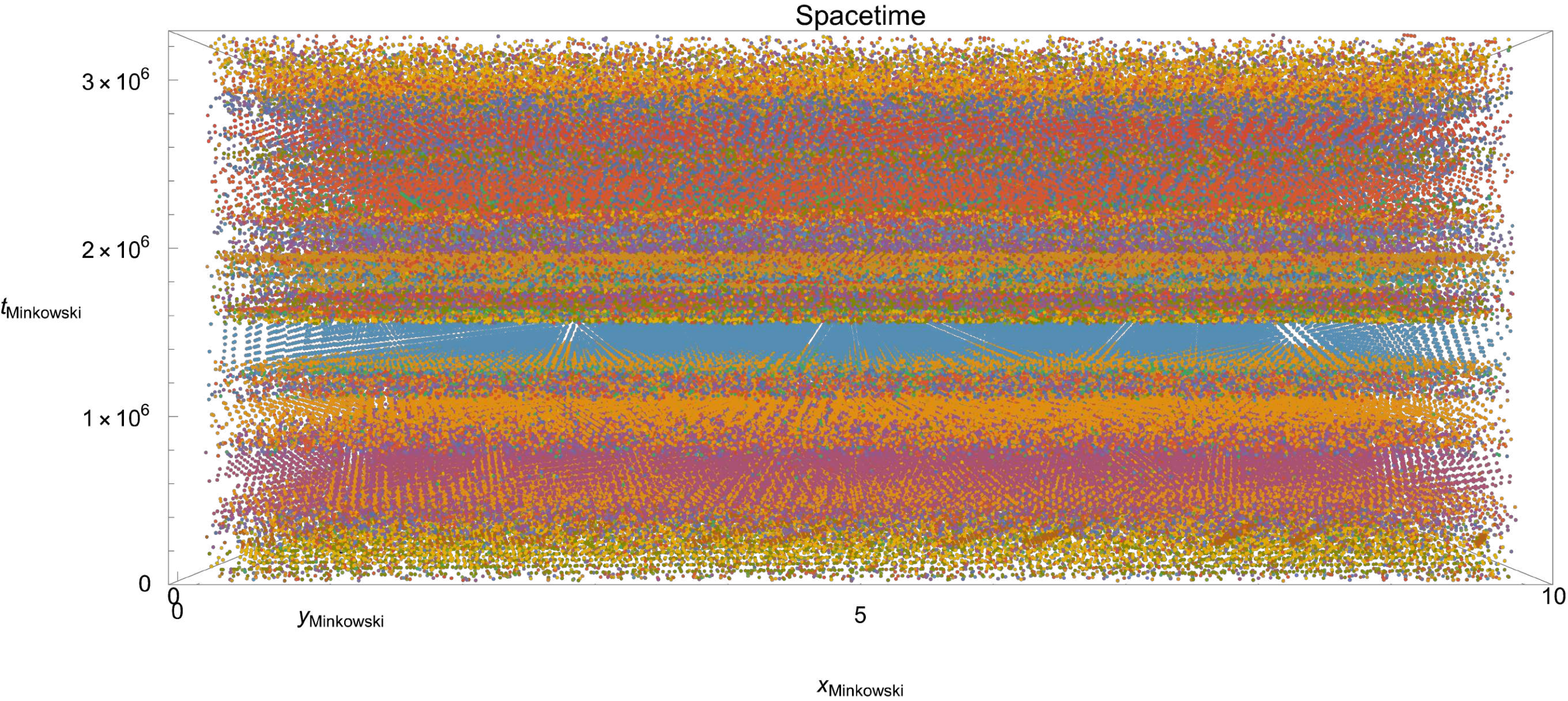} 	\caption{An irreversible run with $N_{\rm familes} = 20$ and $N_{\rm events} = 5 \times 10^5$. The pictorial overload of the half million simulation points is necessary in this figure for density illustration purposes, as well as later examination in further detail, which we perform in Fig.~\ref{fig:main_result_time_slices}. Parameters: $L=10$, $\bar{\sigma}=0.07$, $dt=\bar{\sigma}$, $t_{{\rm max}}= 2\times L$, $\Omega=0.001$. The breaking and reforming of crystal structures is our main result, each crystal being the settling of the system into one of its limit cycles. The random events break the system out of a limit cycle, to later settle into another. This phenomena is due to discausality, a form of retrocausality already found in the 1d+1 case~\cite{realismII}. }
	\label{fig:main_result}
\end{figure*}

\begin{figure*}%[t]
	\centering
$\begin{array}{cc}		
	\includegraphics[width=85mm, height=55mm]{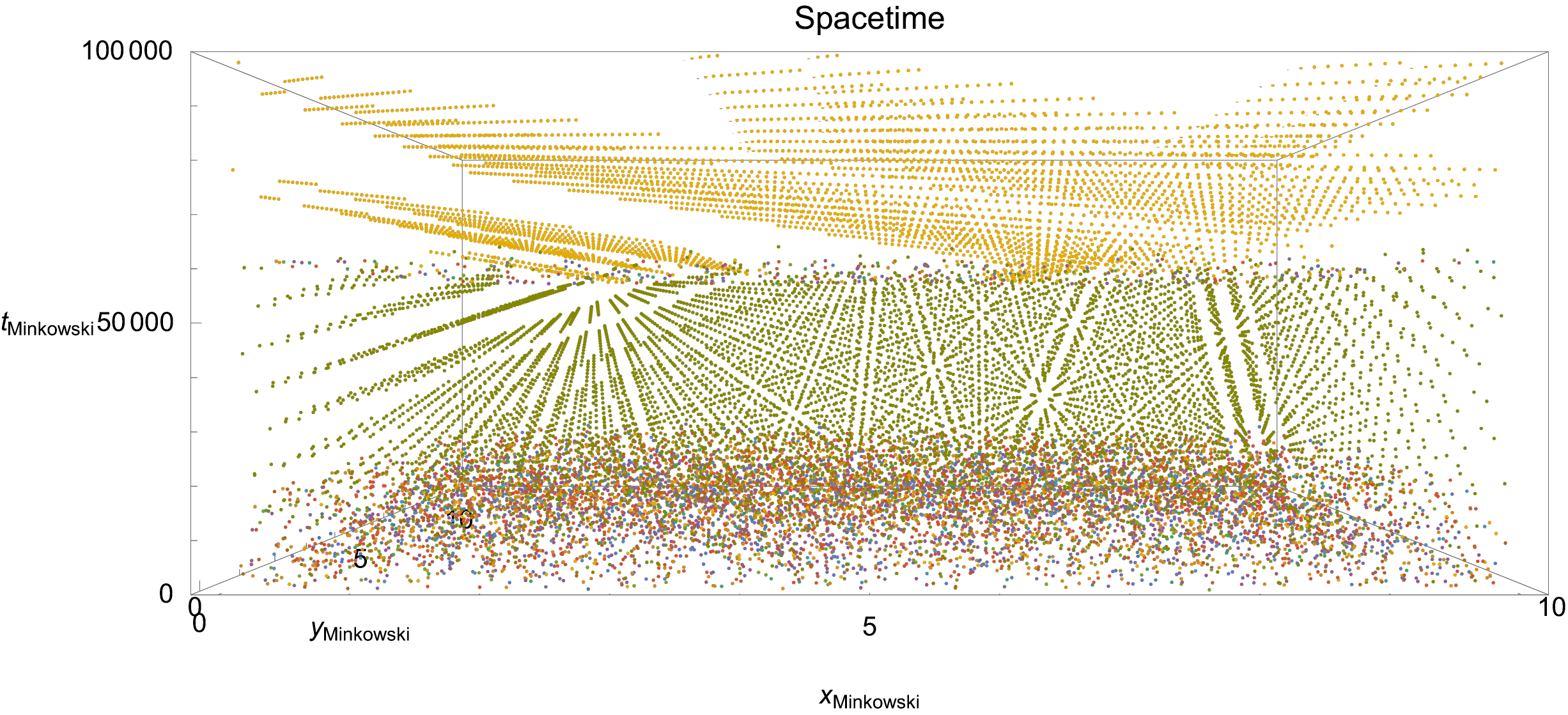} \quad &\quad 
	\includegraphics[width=85mm, height=55mm]{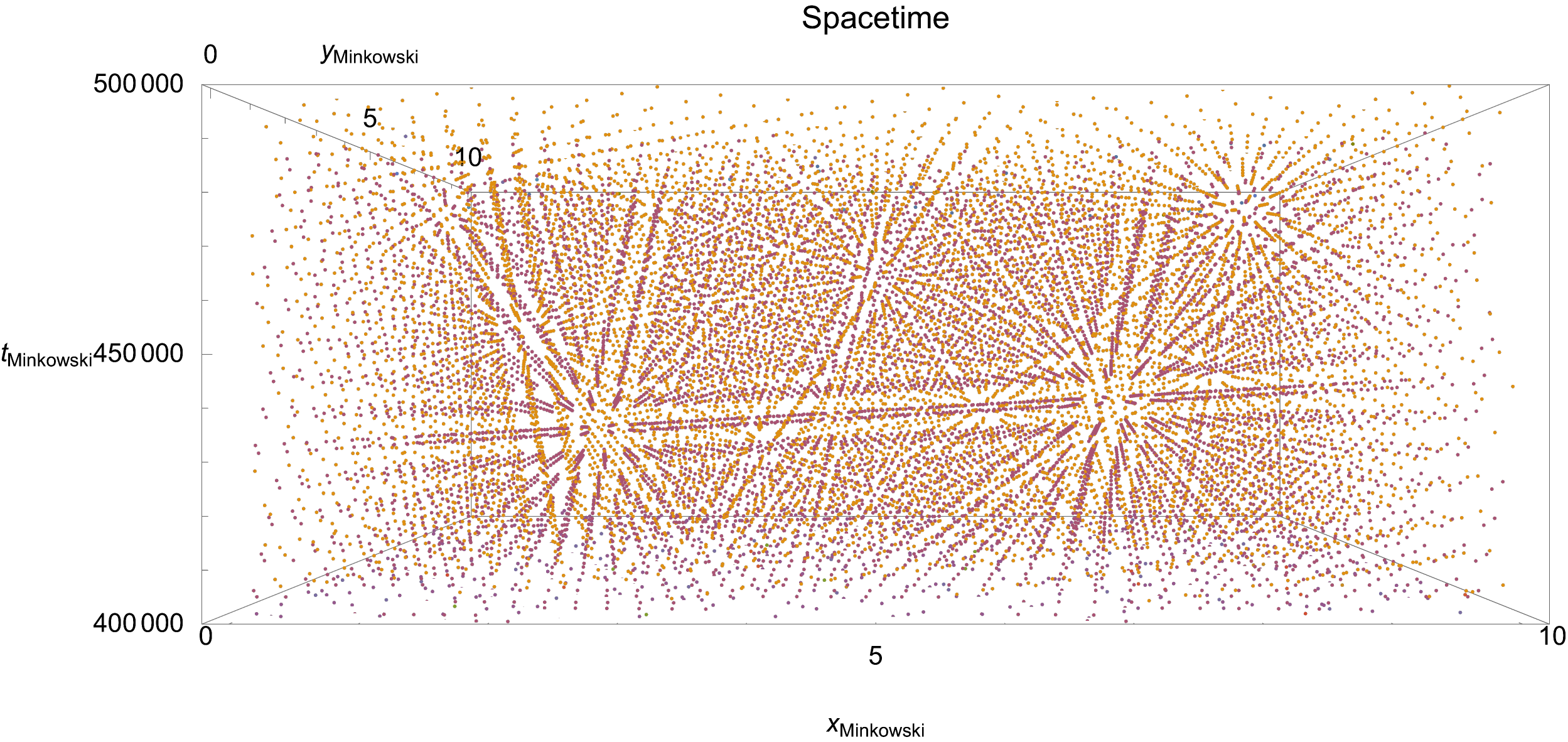}\\
		\textbf{(a)} & \textbf{(b)} \\
		~\\
	\includegraphics[width=85mm, height=55mm]{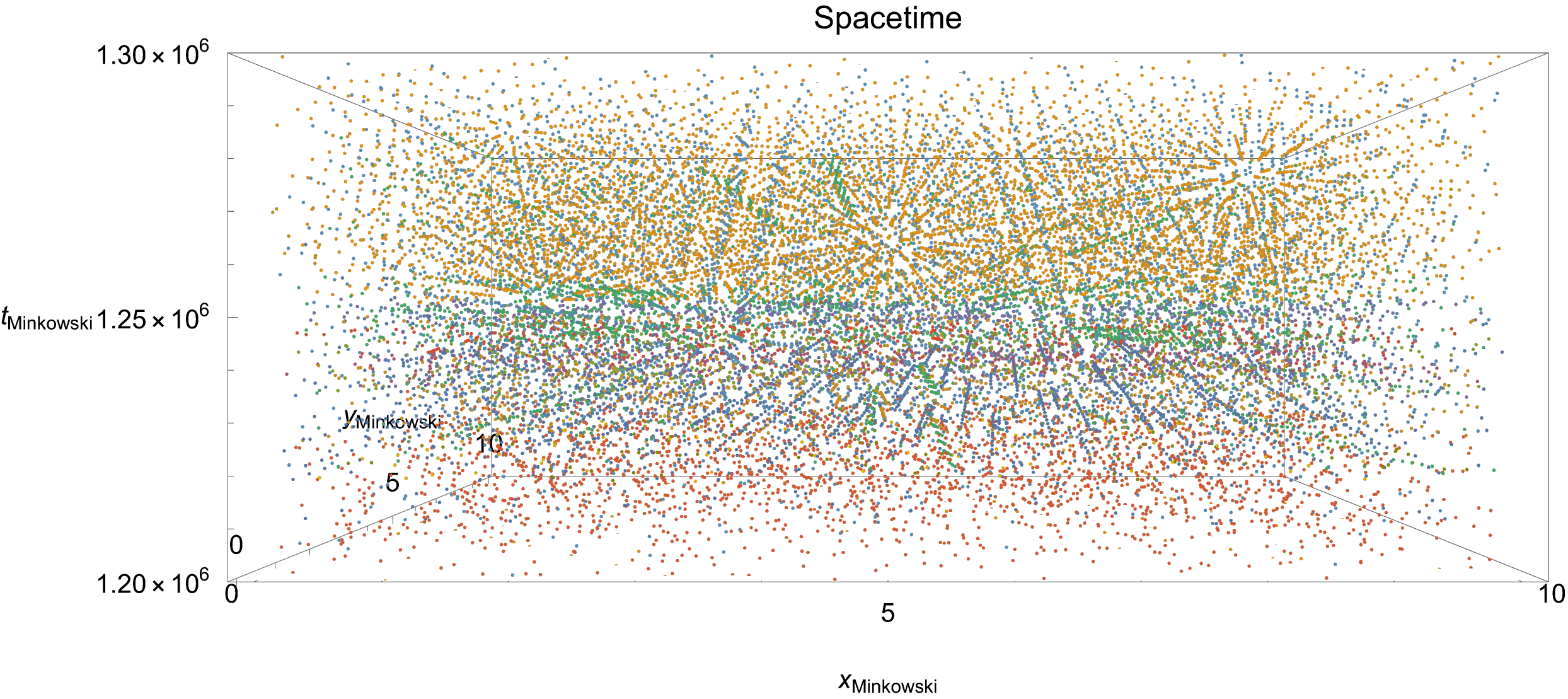} \quad &\quad 
	\includegraphics[width=85mm, height=55mm]{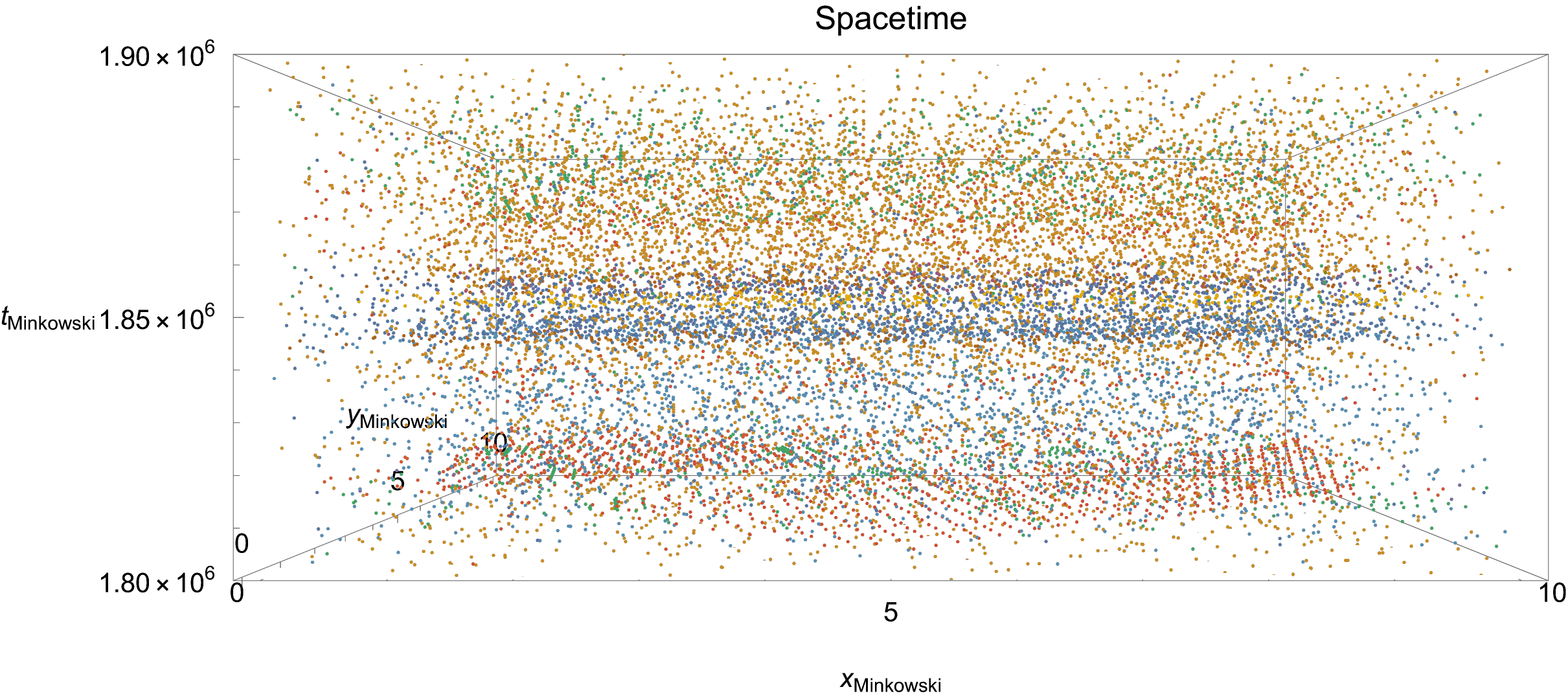}\\
		\textbf{(c)} & \textbf{(d)} \\
		~\\
		\end{array}$
	\caption{Four time slices of our main result from Fig.~\ref{fig:main_result}, with $N_{\rm events} = 5 \times 10^5$, $N_{\rm families}= 20$, and $\Omega = 0.001$. Plot \textbf{(a)}, from 0 to $10^5$ time units shows the disruption of a crystal followed by a small period of chaotic behaviour before the formation of a new crystal. Plot \textbf{(b)} from $4 \times 10^5$ to $5 \times 10^5$ time units shows two superposed crystals. These weren't created at the same time, meaning the run first created one and only later created the other; it didn't alternate between the creation of the two. This is discausality at work. Plots \textbf{(c)} \& \textbf{(d)} show a variety of things, including superposed crystals, crystals superposed with chaotic behaviour which is again due to discausality, and just chaotic behaviour.  }
	\label{fig:main_result_time_slices} 
\end{figure*}

The way we introduce randomness in 2d+1 is by discontinuously disrupting the dynamics. This forces the system to leave its limit cycle and enter a new one, which can be observed by the formation of a crystal of a different color. Also the discausal jumps observed in the companion paper~\cite{ECS2d2} are a further confirmation of this behaviour already observed in the 1d+1 case. 

We also suspect that the lack of formation of several crystals at the same time without randomness might be due not just to how much more difficult it is to enter the time-symmetric phase in 2d+1, but also to the absence of constant novelty being injected. When we introduce randomness to the system it is common, given the run is long enough and there are enough families, to observe lots of simultaneous crystals.This does not constitute a compromise for our results since we still observe several limit cycles forming which means that the system is always able to overcome the input randomness. Moreover many times these are observed to happen at the same Minkowski time but not simultaneously when it comes to the event generator, hence the discausal jumps, a behaviour similar to the 1d+1 case. 

Figure~\ref{fig:main_result} shows this behaviour for a run with 20 families, $5\times10^5$ generated events and 0.001 randomness. In the companion paper~\cite{ECS2d2} we will show a run with more randomness but increasing the randomness just leads to more ill-defined crystals to the point where no structure is visible. Furthermore if we take the randomness to be $1$,  meaning all the parents are randomly selected, the distribution of events amongst all families approaches that of the theoretical relative frequency, ({\mbox{Number of events}$/$\mbox{Number of families}). 

Figure~\ref{fig:main_result} shows the full run, while Fig.~\ref{fig:main_result_time_slices} show portions of $t_{\rm Minkowski} = 10^5$ so that some of the structures involved can be seen more clearly. The system enters and leaves different limit cycles, represented by the formation and destruction of crystals separated be periods of chaotic behaviour lacking a pattern of space-time positions which is the manifestation of the time-asymmetric phase.

We will not go into a further study of this topic in this paper since the purpose here is to present the 2d+1 results. A further detailed study can be found in a companion paper to this one~\cite{ECS2d2}. In the 1d+1 case randomness was injected constantly and in a more subtle manner compared to the discontinuous and abrupt one here, but both lead to the same end result of forcing the system out of a limit cycle. In the 1d+1 case there are in general a few more families forming the  trajectories. This is because even though we increased one dimension, we are still using around the same number of families as in the 1d+1 case for computational reasons. Ideally the number should be increased more dramatically to enhance the likelihood of more suitable pairs of rays forming a crystal. For instance, if we use 40 initial families in 1d+1 we should probably use over 200 in 2d+1, which is a fraction of $40^2$, to see more families forming the crystal and observe more variety of structures. 

\section{Robustness tests}

For robustness we show two runs of the {\bf same two main sets of parameters} with different initial momenta, for the deterministic and irreversible runs.

\begin{figure*}%[t]
	\centering
$\begin{array}{cc}		
	\includegraphics[width=0.8\textwidth]{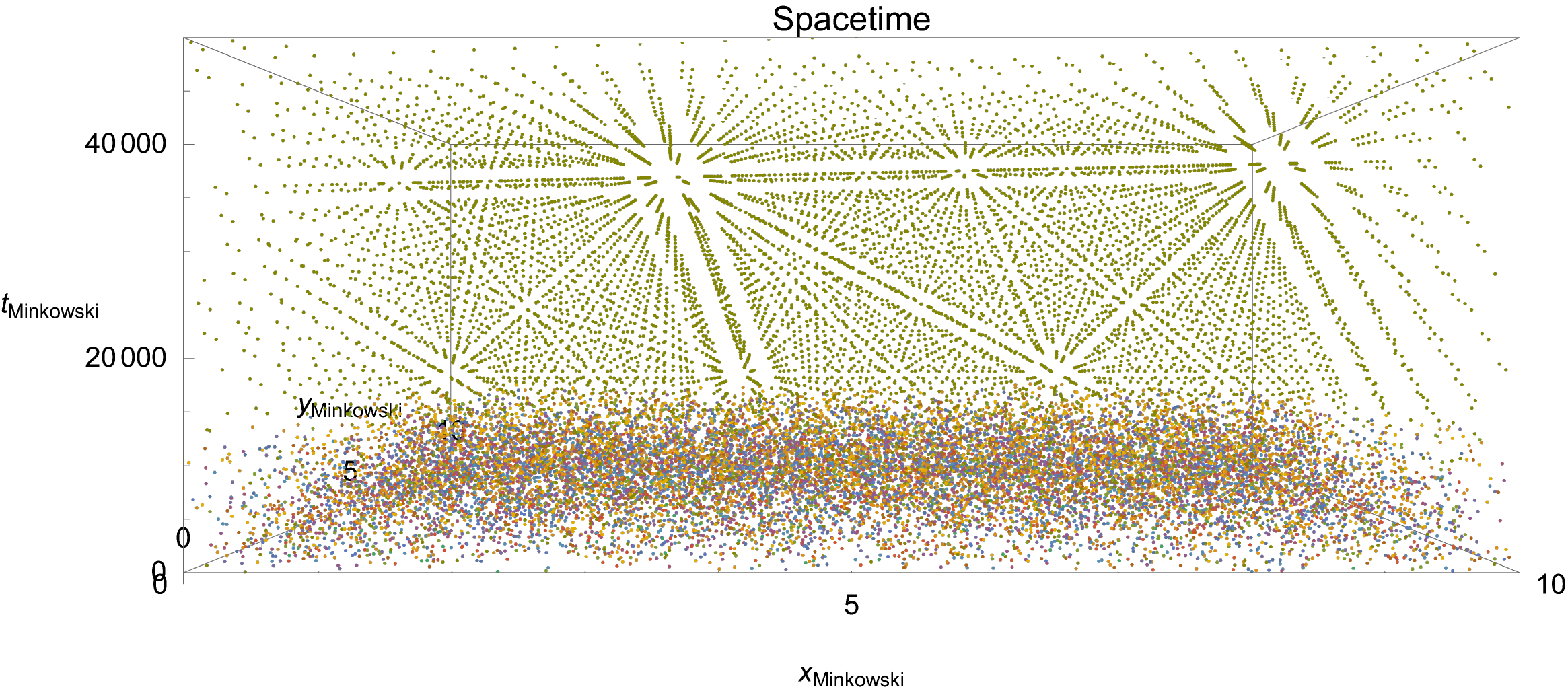}\\
    \textbf{(a)} \\
    ~\\
	\includegraphics[width=0.8\textwidth]{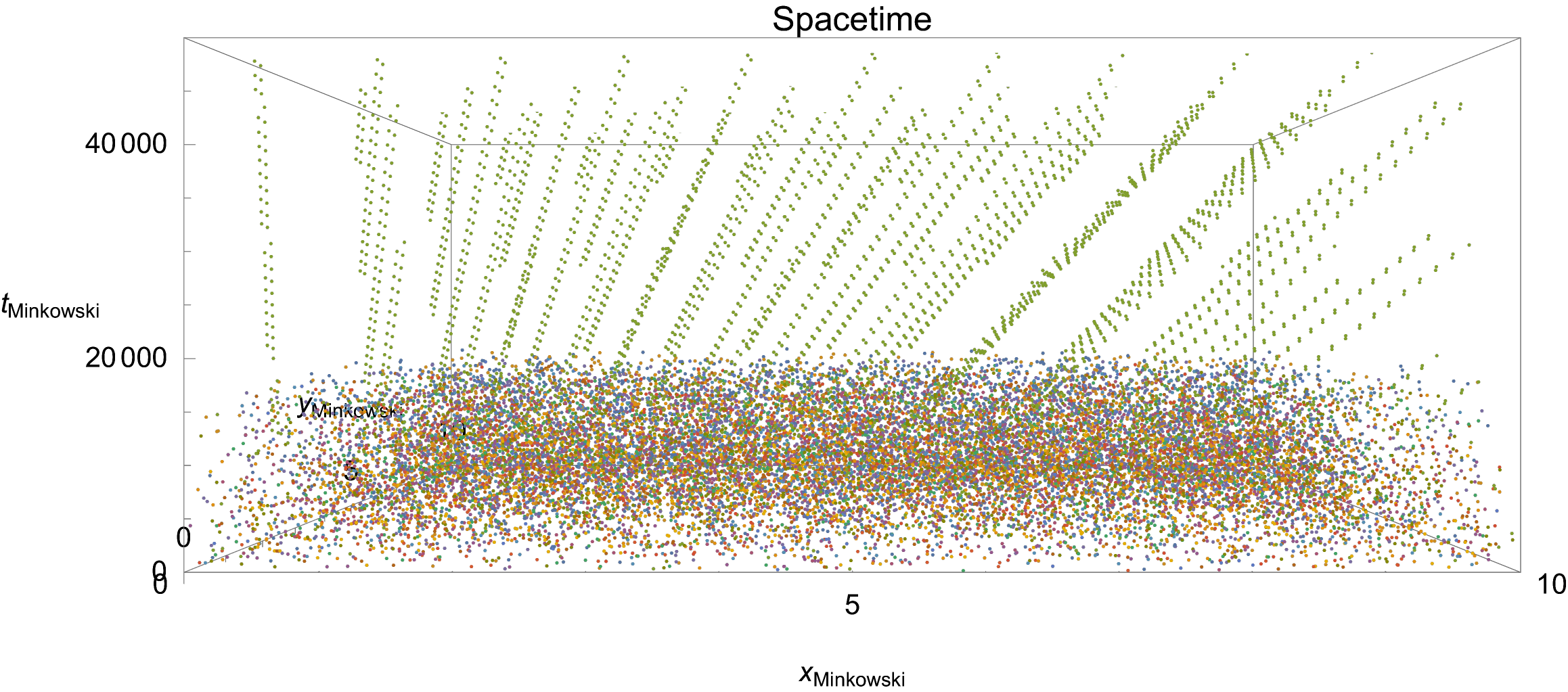}\\
    \textbf{(b)} \\
    ~\\
	\end{array}$
	\caption{Two deterministic runs with the same parameters to convey variations between the runs due to different initial conditions. Parameters: $L=10$, $dt=\bar{\sigma}$, $t_{{\rm max}}=2\times L$, $\bar{\sigma}=0.07$, $N_{\rm  families} = 40$, and $N_{\rm events} = 5\times 10^4$. 
 We observe that the crystal structures are different, due to the orientation of the rays forming the structures.}
	\label{fig:deterministic_comparison} 
\end{figure*}

\subsection{Deterministic case}

Figure~\ref{fig:deterministic_comparison} shows two runs with the same parameters of deterministic runs. The fact that the crystals are roughly the same color and emerge around the same Minkowski time is a mere coincidence. The difference between them however is clear; structurally these are very different crystals, a reflection of the different momenta involved in the interactions.

\subsection{Irreversible case}

Figure~\ref{fig:random_comparison} shows two random runs, again with the exact same parameters. The fact that the crystals are different is not the main fact to observe here. The biggest difference in the case is how and when the system enters and leaves its limit cycles, which equates to the formation and breaking of the crystal respectively. These are very different and now, they are not only due to the momenta involved which are different to both runs, but also due to the events that were generated with random parents. There is the same probability in both runs, but naturally events occur at different times and can have a different impact of each run since the momenta involved are different.

\begin{figure*}%[t]
	\centering
$\begin{array}{cc}		
	\includegraphics[width=0.8\textwidth]{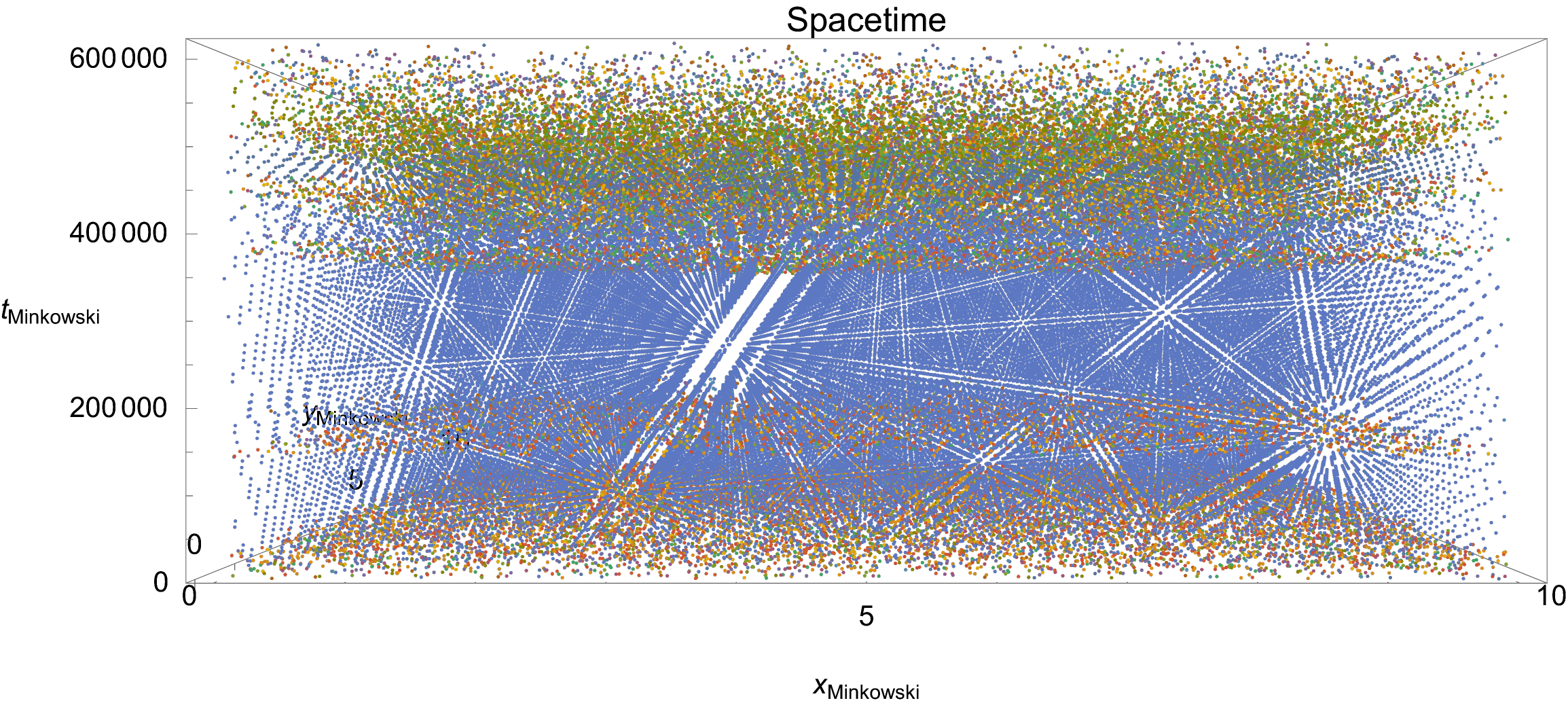}\\
    \textbf{(a)} \\
    ~\\
	\includegraphics[width=0.8\textwidth]{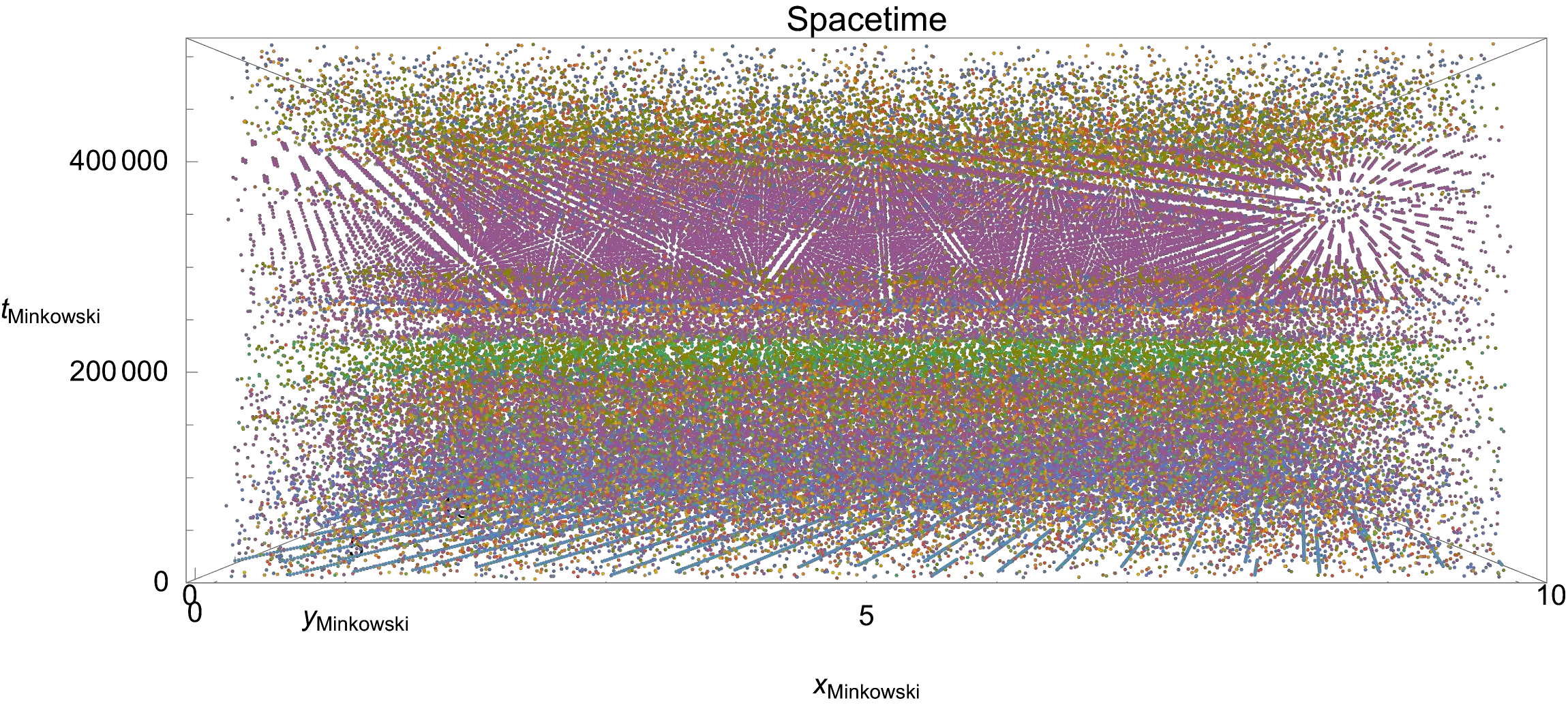}\\
    \textbf{(b)} \\
    ~\\
	\end{array}$
	\caption{Two plots with the same parameters, to convey the differences between randomness due to the initial conditions but now with stochasticity which is the same for both plots. Parameters: $L=10$, $dt=\bar{\sigma}$, $t_{{\rm max}}=2\times L$, $\bar{\sigma}=0.07$, $N_{\rm families} = 20$, $N_{\rm events} = 10^5$, and $\Omega = 0.001$. The main difference to observe here is the formation and breaking of the crystals and how many families can form them. Plot \textbf{(b)} shows more variety in this regard, with stable trajectories in blue very close to the beginning of the run. Moreover the formation and breaking of the crystals is highly influenced by the events generated with random parents and these will naturally happen at different times with varying impact on the runs. }
	\label{fig:random_comparison} 
\end{figure*}

\section{Review of the energetic causal sets programme}

\subsection{Historical background}

The energetic causal sets programme is a quantum gravity approach which started in 2013 \cite{ecs}, motivated by the fundamental principle of causality and the irreversibility of time as the pinnacles that bring forth, and support,
foundational theories of Nature. We proposed a model of the Universe in which every event is unique and part of a global network of causal connections with other events. This uniqueness of events, evidenced by their unique position in the network structure, also gives each event a unique view and perspective of the global network from that event's node.

In the initial paper \cite{ecs}, Cort\^es and Smolin derived the emergence of classical, manifestly time-symmetric Lorentzian manifolds from the underpinning causal network. The second work \cite{quantum_ecs} is the quantum formulation of the classical model of Ref.~\cite{ecs}. There the path was laid down for the emergence of quantum mechanics from the same causal network structure, which we called {\it Quantum Energetic Causal Sets}. Several aspects of the quantum theory were shown to arise straightforwardly from the existing causal setup, though not all as we now describe. For example, since space-time does not exist in the underpinning regime for ECS, locality or non-locality cannot be claimed as properties or, conversely, as absent properties in the ECS quantum gravity program. {\it Causality} is the only surviving principle in the fundamental regime. 
%$\hbar$ \andrew{Why is $\hbar$ written here?}

Following that work, we showed in Ref.~\cite{spinfoams} that energetic causal sets have a one-to-one correspondence with a class of spinfoam models proposed by Wieland in Ref.~\cite{wieland}. This implies that all results of the successful spinfoam models programme can be readily imported into energetic causal set model studies. 

Cort\^es and Smolin then discovered a novel interpretation of the phase transition observed in ECS models \cite{limitcycle}. The  transition from the time-asymmetric to the time-symmetric regime is an example of system-capture by limit cycles, a well-known phenomenon of cellular automata, and more generally of discrete dynamical systems. This discovery provided a measure of how ubiquitous such capturing and trapping of  dynamics is in Nature, endowing the ECS phase transition with a framework which requires very few conditions to occur. 

The most recent step in the programme consisted of a close examination of quantum ordering of events. There exist claims of violations of causality in a particular interpretation of quantum mechanics, the Aharanov interpretation, namely the Two-State Vector Formalism (TSVF) \cite{TSVF}. In TSVF this violation of causality is called {\it retrocausality}. We investigated whether such violations of causality can be observed in energetic causal sets as well, and found that, although the sequence of network events does not always follow the manifold ordering, the generation of events always obeys the causality principle and the causal (partial) ordering of the network is not violated \cite{realismI,realismII}. Instead of partial order we can also say {\it causally-thick order}.

%%%%%%%%%%%%%%%%%%%%%%%%%%%%%
%%%%%%%%%%%%%%%%%%%%%%%%%%%%%
\subsection{The path forward: from irreversibility to non-reductionism}

%\andrew{Changed from section to subsection of Section VII, but not certain that was what you intended.}

The irreversibility line of enquiry has led our collaboration to question similarly innocuous principles which are so universally adopted in physics that one would not easily reconsider their validity. One such principle is that of reductionism, on which the majority of the theoretical physics body of knowledge, as we know it today, is grounded. Reductionism is tightly connected to time-symmetric trajectories in the following way. In the reductionist assumption we assert that the world can ultimately be broken down into the most fundamental individual constituents. Following the same assumption we can equally build up the world again from those same elementary constituents. This implies that such conversions or translations from the macroscopic world to the microscopic world and vice-versa needs to be governed by time-symmetric equations. If the equations enabling such zooming in and out of the microscopic description were not time-reversible, then the translation between the very small and the very large would not be reproducible. Or, at the very least, we would arrive at a different microscopic description of the elementary world since we would be unable to invert and retrace the equations back to the same fundamental regime. 

A long reflection on the limitations of the reductionist assumption, and hypothesizing a possible breakdown of reductionism, enabled two of us, along with Kauffman and Smolin, to propose a quantitative translation between phase spaces in physics and those in living systems. We proposed the new discipline of biocosmology \cite{biocosm}.

In recent years the reductionist assumption has been considered and questioned by various members of our theoretical and phenomenological physics community \cite{anderson,leggett,laughlin,wiltshire,wmid,Palmer}. In Ref.~\cite{Hertog} Hertog reflects at how, later in his life, Stephen Hawking considered that non-reductionist, top-down approaches could be necessary to complement bottom-up approaches for the completion of physical picture of the world. Hertog gives the AdS/CFT duality as an example, which is however not the analogy we have followed in the analytical setup for biocosmology (where we have considered tools of combinatorial innovation).

%%%%%%%%%%%%%%%%%%%%%%%%%%
\section{Conclusions}
This article constitutes a significant milestone in the ECS quantum gravity programme because we surpass the obstacle of extending the 1d+1 initial work to higher dimensions.
We have done so for 2+1 dimensions, building on the seminal 1d+1 algorithm introduced in Ref.~\cite{ecs} and subsequently developed in Refs.~\cite{quantum_ecs,spinfoams,limitcycle,realismI,realismII,margoni,trinity,topos,leeint}. 

The principal new requirement in moving to higher dimensions is that the particles mediating event creation need to have a finite extent, or cross-section, to enable interactions. This then necessitates care in defining the location of newly-created events. We described how our choices are motivated, guided by the 1d+1 set-up, and described the various algorithmic parameters necessary for implementation.

Our simulations show the same transition to an effectively time-symmetric phase that was discovered in 1d+1~\cite{ecs}, in this case giving the visual appearance of a crystal corresponding to the establishment of stable quasi-particles. In a run whose event generator is entirely deterministic, the crystals are persistent. We also show that the addition of a trace amount of stochasticity to the algorithm is sufficient to disrupt the time-symmetric phase, with a new limit cycle becoming established after a further period of time. This breaking and reforming of the effective time-symmetric structure under randomness is the main result of our study.

This article defines our 2d+1 algorithm and illustrates our main results. A companion paper \cite{ECS2d2} makes a detailed exploration of the effect of varying the model parameters, along with an investigation of the causal network of the events. Future work will extend the analysis to the realistic case of 3+1 dimensions. We do not expect this to raise any new points of principle beyond the case investigated here, merely additional computational complexity and demands. Establishing the 3d+1 case will be a key milestone for the ECS programme.

The work presented also includes a very promising outlook on the examination of fundamental causality in the underlying network description of the 2d+1 model, which we are currently in the process of detailing. On top of that we have hinted at a methodology for the inclusion of one further spatial dimension, which, when carried out, will fully complete the emergence of classical space-time manifolds from the ECS proposal of quantum gravity. Recall that a prescription for the inclusion of spatial curvature is straightforward in ECS, already provided in Ref.~\cite{ecs}. At the same time, as we gear up to tackle the question of curvature in quantum mechanics  it is interesting to note that there is no unifying understanding of this concept in the context of quantum gravity \cite{qgconf}.

\section{The role of science as unifier of threads: Quantum Gravity, Biocosmology, and Artificial General Intelligence}

\label{sec:IX}

The completion of the manuscript is fortuitously timed because it informs an intense debate taking place as we write in the field of computer science. Our work establishes a link between fundamental physics and the Artificial General Intelligence (AGI) debate in the following sense: our programme of conceptualising the Universe as model of unique events proposes a model of reality in which every event in the observable universe is a fundamental and essential constituent of the reality we observe.
% NEXT PHRASE MEANING UNCLEAR, and try to describe in studies of quantum gravity in the theoretical physics community. 
A key property of the model is that all events are unique and distinguishable from each other. In the ECS sense reality cannot be compressed, nor summarised in an information theoretical sense, without loss of key correlations in the causal network. In turn these lead to loss of entanglement information which occurs during the severing of causal links that takes place when forming the compressed subsystem which is to be a proxy for reality. 

In a quantum mechanical description one refers to such a severed subsystem as a mixed state. This means that the associated Hilbert space lacks the information to recover a complete, coherent description of the quantum world. This severing often leads to conceptual inconsistencies like the measurement problem and violation of Bell inequalities, as well as the incompleteness of quantum information \cite{cohen-tanoudji,bell,aspect,clauser,zeilinger}.

%and the black hole information  paradox 
%\cite{statisticalindependence_counterfactuals_informationparadox}  

The computer scientist and founding father of virtual reality Jaron Lanier (Microsoft Research) recently referred to the same concept of the incompressibility of reality. Lanier asserts that compression of a physical state reality cannot occur without associated loss of information. This means that the envisaged reconstruction of the real state from the compressed information in the pseudo-state, will not recover the original physical state. All bits of information count, and are necessary, if one wishes to accurately describe  physical reality which, after all, is the object of study that we theoretical physicists and quantum gravity theorists devote our lifetime to. One cannot drop bits of information without inevitable loss of accuracy in the recovered model. Lanier refers to this property in Ref.~\cite{JaronNY2} as ``Reality is irrepressible''. 

A direct implication of Lanier's thinking, which the technologist and computer philosopher has defended for decades, is the view that artificial general intelligence is not attainable \cite{JaronNY1}. One would need all the individual unique events (in the causal network) constituting the intelligent living system to replicate it in the digital world. This conceptual framework and vision of reality finds a direct parallel in the model of reality as a process of unique events, as advocated by us in the ECS programme. 

In this way the ECS programme of quantum gravity provides a vision, and establishes a direct parallel to the AGI debate. ECS are another way of stating that physical reality is not to be replicated in digital information, with the inevitable consequence that human-like AGI is equally impossible to attain in digital format.

Lee Smolin wrote in Ref.~\cite{timereborn} : ``Thus, I have observed that those who believe in strong artificial intelligence (AI), tend also to believe that nature is a computation, in the illusory character [...] of the passage of time, in the many worlds interpretation of quantum theory as well as in anthropic multi-verse cosmologies.'' He also wrote: ``Persons have properties that cannot be captured by any computation. The brain is a physical system, but not a programmable digital computer.'' In the biocosmology programme, which follows up from the initial proposition of time irreversibility, we developed the mathematical framework which provides a quantitative demonstration of this assertion \cite{biocosm}, 

And so, in this section we have lined up a wide suite of arguments, enabling us to establish a unifying way of thinking and a framework connecting quantum gravity, biology (and biocosmology), and computer science. All three come together to articulate and fortify the argument that human intelligence will not find its match in  digital technology.  

%emulated.

\begin{acknowledgments}
We thank Catarina Ferreira and Lee Smolin for discussions. This research was supported in part by Perimeter Institute for Theoretical Physics. Research at Perimeter Institute is supported by the Government of Canada through Industry Canada and by the Province of Ontario through the Ministry of Research and Innovation. This research was also partly supported by grants from NSERC and FQXi. This work was supported by the Funda\c{c}\~{a}o para a Ci\^encia e a Tecnologia (FCT) through the research grants UIDB/04434/2020 and UIDP/04434/2020. M.C.\ acknowledges support from the FCT through grant SFRH/BPD/111010/2015 and the Investigador FCT Contract No.\ CEECIND/02581/2018 and POPH/FSE (EC). A.R.L.\ acknowledges support from the FCT through the Investigador FCT Contract No.\ CEECIND/02854/2017 and POPH/FSE (EC). M.C.\ and A.R.L.\ are supported by the FCT through the research project EXPL/FIS-AST/1418/2021. 
\end{acknowledgments}

%\clearpage

\appendix*

\section*{Appendix: Parameter control, computational parameters, and system tuning}

%\addcontentsline{toc}{section}{Appendix: Parameter control, computational parameters, and system tuning}

\subsection{Parameter control}

%\marina{Shortening this section considerably, study of families 4 plots, and 2 figures $\bar{\sigma}=0.02, 0.05$ plots as is.}

Here we describe how the parameters controlling our system influence the emergence of the time-symmetric phase. We are dealing with a chaotic and stochastic system so two runs with the same parameters will typically yield different results. However there are patterns in the outcomes of the runs when we change our parameters. It is important not only to understand the effect of each parameter on the runs but how their effects interact. For example, if we decrease the particle radius then the maximum time for interaction must be increased in order to obtain similar results. 

To best see the effect of changing the various parameters on the emergence of the crystal, we first eliminate the randomness in pair interaction choice so the effects are just due to the change in the parameters.  We only use the deterministic event generator for this discussion. 

\subsection{Search time and time step: $t_{{\rm max}}$ and $dt$}

The parameter $dt$ must be significantly smaller than the radius of a particle or interactions will be missed. Ideally $dt$ should be as small as possible while using computational power efficiently. Our experience with the simulations tells us that setting $dt=\bar\sigma L/10$ ensures accurate collision identification without excessive computational time. This makes $dt$ always one tenth of the radius of each ray. The effect of this parameter is not palpable in plots, so no figures are presented.

\begin{figure*}%[t]
	\centering
$\begin{array}{cc}		
	\includegraphics[width=80mm]{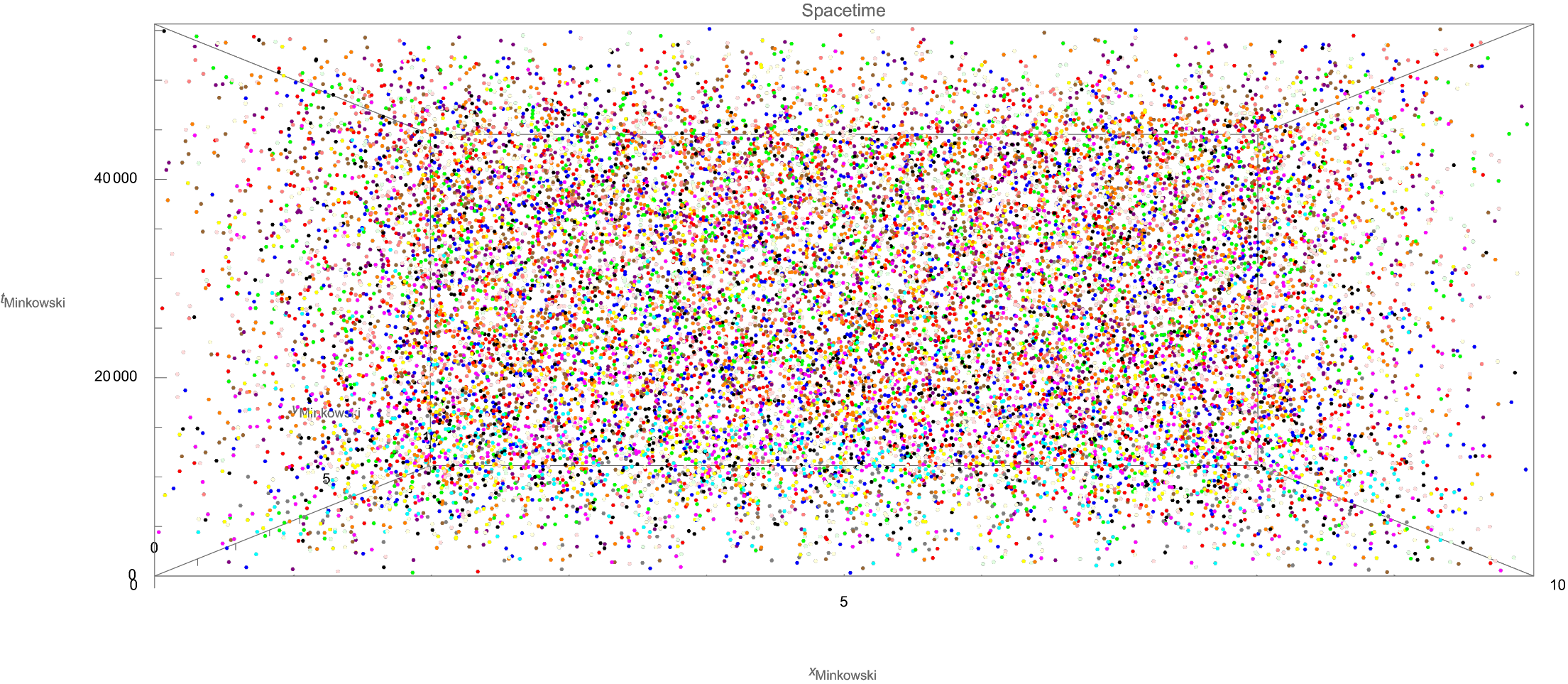} \quad &\quad 
	\includegraphics[width=80mm]{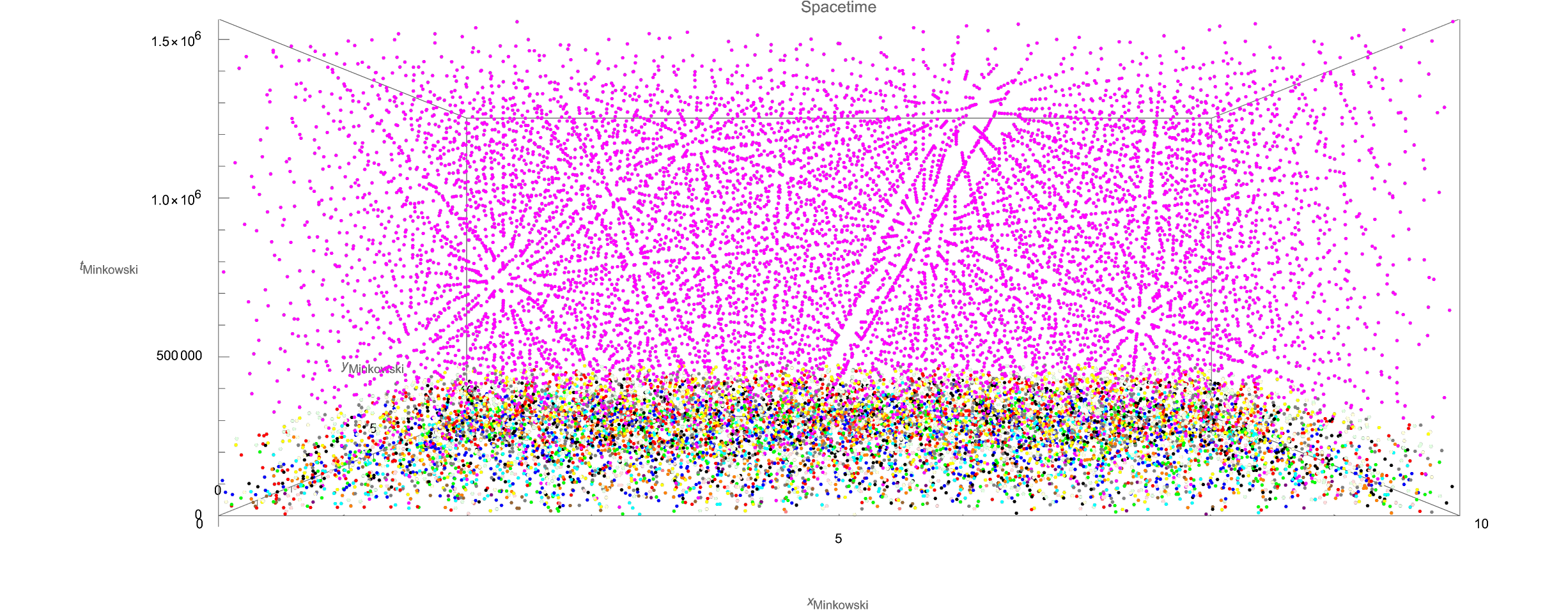}\\
		\textbf{(a)} & \textbf{(b)} \\
		~\\
		\end{array}$
	\caption{The effect of the variable $t_{{\rm max}}$ on runs with zero randomness. Parameters: $\bar{\sigma}=0.025$, $dt=\bar{\sigma}$, $L=10$, $N_{\rm families} = 15$, $N_{\rm events} = 2 \times 10^4$. Plot \textbf{(a)} shows a run with $t_{{\rm max}}=1\times L$ and Plot \textbf{(b)}  $t_{{\rm max}}=100\times L$. The importance of this parameter lies in it being a limiting factor on the emergence of the time-symmetric phase.
 Because a very small value, depending on other parameters like $\bar\sigma$, can make our rule obsolete since it may force very distant pasts to be the ones chosen for interaction. This leads to an inconsistent application of the dynamical rule, not allowing the time-symmetric phase to emerge.}
	\label{comparetmax} 
\end{figure*}

The time we wait to see if two chosen particles meet before switching to a different pair is denoted $t_{{\rm max}}$, counted from the time of the younger parent. Compared with the initial number of families, this variable's behaviour is harder to analyze since its effect on the outcome of a run is strongly dependent on other parameters. So in order to show its effect we need to fix the other variables first.

Recall that $t_{\rm max}$ should be significantly greater than the mean time to interaction given in Eq.~(\ref{e:tbar}), which (in units of the box size) depends only on the cross-section $\bar\sigma$. There is an additional effect from the number of families, that if this is small we may run out of pairs to try if $t_{\rm max}$ is set too small. Let's now see how $t_{\rm max}$ is related to the number of families and $\bar\sigma$.

Firstly if we have a small $t_{{\rm max}}$, a small $\bar\sigma$, and a small number of families, it is  likely not to be possible to find two particles that collide in the available $t_{{\rm max}}$, forcing the run to terminate. This can be solved by increasing the number of particles present, but this raises another problem. Every time the selected particles don't interact in the available $t_{{\rm max}}$, the two particles with next closest pasts are selected. If $t_{{\rm max}}$ is too small for the selected cross-section many pairs will be tried and the chances are, by the time a pair is found that can interact in the selected $t_{{\rm max}}$, the difference in pasts of this pair is far from the minimum value. This makes the rule obsolete not allowing for the emergence of a time-symmetric phase even without any randomness being injected after the beginning of the run, and is why we should seek to correlate $t_{{\rm max}}$ to $\bar\sigma$. To show this effect we tune our system in the following way: We want the emergence to be hard so we need our cross-section $\bar\sigma$ to be small, a big number of families and, given these choices, a lengthier run.

Figure~\ref{comparetmax} shows two plots describing the behaviour we discussed with $N_{\rm events} = 20\times 10^4$, $N_{\rm families} = 15$, and $\bar\sigma$ of $0.025$, giving a mean time to interaction of $\bar{t} \simeq 7 L$. Plot \textbf{(a)} shows a run with a $t_{{\rm max}} = 1\times L$ and Plot \textbf{(b)} a run with $t_{{\rm max}} = 100L$. In Plot \textbf{(a)} there is no emergence of a time-symmetric phase whereas in Plot \textbf{(b)} a clear crystal emerges. This is because in the former case the event generator is trying many pairs until it finds one that interact in the given time. This doesn't lead to a consistent application of the dynamical rule.

Another effect, which is irrelevant for our purposes but mentioned for sake of completeness, is that a bigger $t_{{\rm max}}$ makes the runs longer in Minkowski time, since we are allowing interactions to take longer Minkowski time to occur. The only effect is that some pasts might distinguish themselves from the rest faster due to a bigger time component of the interactions. This however, does not translate directly into a faster or slower transition to the time-symmetric phase.

\subsection{Tuning the system}

Of all parameters only $N_{\rm families}$ and $\bar\sigma$ are physical, the others are computational. We would like to be able to vary the physical parameters without compromising the results because of the computational ones, meaning that the results are a consequence of the true dynamics of the system, not the computation.

Tuning the system to obtain good results requires first that we get a feel for the effect of the parameters. They are all interconnected and moreover we are dealing with a chaotic system, highly influenced by the initial conditions. Many tests were needed to optimize the parameters. The choice we found had the following parameters: $dt=\bar\sigma$, $\bar\sigma=0.1$, $t_{{\rm max}}=1\times L$.

The number of families only affects the speed of convergence to the time-symmetric phase. 
We would however like the fixed number of initial events of the reference run to relate $t_{{\rm max}}$ to $\bar\sigma$. Given the reference run, for one unit of radius it takes on average one winding for an interaction to occur otherwise the crystal wouldn't form consistently. If we increase the area by increasing the radius we should expect a proportional decrease in windings and vice-versa. 

Finally, how much randomness we should put in the system? This depends on the parameters chosen, which is why we compare to a reference run; if we change the computational parameters we can change the radius, which is the real parameter, whereas $t_{\rm max}$ is merely a computational parameter, leading to the same outcomes. The reference run shows that it takes on average 1000 events for a crystal to form for 15 families. So let us think of randomness in terms of 1 random event per 1000 created. We will see in Ref.~\cite{ECS2d2} that one event can be enough to disrupt a crystal. We already took $dt=\bar\sigma$ so we effectively related all the computational parameters to all physical parameters. As the randomness and the number of events are physical parameters, it is best to vary them independently.

\end{document}